\documentclass[aps,prb,twocolumn,showpacs,preprintnumbers,prb,superscriptaddress]{revtex4-1}
\usepackage{graphicx}  
\usepackage{dcolumn}   
\usepackage{bm}        
\usepackage{amssymb}   
\usepackage{framed}
\usepackage{amsmath}
\usepackage{hhline}
\usepackage{subfigure,amsmath,verbatim,moreverb}
\usepackage{tabularx}
\usepackage{lipsum}
\usepackage{longtable}
\usepackage{booktabs}
\usepackage{threeparttable}
\usepackage{booktabs,amssymb,siunitx}
\usepackage{latexsym}
\usepackage{dcolumn}
\usepackage{amsmath}
\usepackage{epsf}
\usepackage{float}
\usepackage{hyperref}
\usepackage{xcolor}
\usepackage[utf8]{inputenc}
\usepackage[normalem]{ulem}
\usepackage{amsmath}
\usepackage{etoolbox}
\newcounter{subeqn} %

\usepackage{mathptmx}

\usepackage{booktabs}
\usepackage{tabularx}
\usepackage{adjustbox}
\usepackage{rotating}
\usepackage{color,soul}
\usepackage{multirow}
\usepackage{relsize}

\usepackage{hyperref}
\hypersetup{
    unicode=true,          
    pdftoolbar=true,        
    pdfmenubar=true,        
    pdffitwindow=false,     
    pdfstartview={FitH},    
    colorlinks=true,       
    linkcolor=blue,          
    citecolor=blue,        
    urlcolor=blue,
}

\begin{document}
\title{Accurate and efficient prediction of the band gaps and optical spectra of chalcopyrite semiconductors from a non-empirical range-separated dielectric-dependent hybrid: Comparison with many-body perturbation theory}
\author{Arghya Ghosh}
\affiliation{Department of Physics, Indian Institute of Technology, Hyderabad, India}
\author{Subrata Jana}
\altaffiliation{Corresponding author: subrata.niser@gmail.com}
\affiliation{Department of Molecular Chemistry and Materials Science
Weizmann Institute of Science,
Rehovoth 76100, Israel}
\author{Dimple Rani}
\affiliation{School of Physical Sciences, National Institute of Science Education and Research, An OCC of Homi Bhabha National Institute, Jatni, India}
\author{Manoar Hossain}
\affiliation{Institut f\"{u}r Physik and IRIS Adlershof, Humboldt-Universit\"{a}t zu Berlin,
Zum Gro$\beta$en Windkanal 2, 12489 Berlin, Germany}
\author{Manish K Niranjan}
\affiliation{Department of Physics, Indian Institute of Technology, Hyderabad, India}

\author{Prasanjit Samal}
\affiliation{School of Physical Sciences, National Institute of Science Education and Research, An OCC of Homi Bhabha National Institute, Jatni, India}

\date{\today}

\begin{abstract}

 The accurate prediction of electronic and optical properties in chalcopyrite semiconductors has been a persistent challenge for density functional theory (DFT) based approaches. Addressing this issue, we demonstrate that very accurate results can be obtained
using a non-empirical screened dielectric-dependent hybrid (DDH) functional. This novel approach showcases its impressive capability to accurately determine band gaps, optical bowing parameters, and optical absorption spectra for chalcopyrite systems.
What sets the screened DDH functional apart is its adeptness in capturing the many-body physics associated with highly localized $d$ electrons. {\textcolor{black}{Notably, the accuracy is comparable to the many-body perturbation based methods (such as $G_0W_0$ or its various approximations for band gaps and Bethe-Salpeter equation (BSE) on the top of the $G_0W_0$ or its various approximations for optical spectra) with less computational
cost, ensuring a more accessible application across various research domains.}}
The present results show the predictive power of the screened DDH functional, pointing toward promising applications where computational efficiency and predictive accuracy are crucial considerations. Overall, the screened DDH functional offers a compelling balance between cost-effectiveness and precision, making it a valuable tool for future endeavors in exploring chalcopyrite semiconductors and beyond.

\end{abstract}

\maketitle

\section{Introduction}
\label{intro}
Over the last two decades, ternary chalcopyrite semiconductors have attracted a great deal of attention due to their applications in renewable and sustainable technologies~\cite{green2019solar,jackson2011new,walsh2012kesterite,feng2011three,rife1977optical,alonso2001optical}. In particular,
these materials are primarily
used as absorbers in thin film solar cells due to their great off-stoichiometric steadiness~\cite{feurer2019efficiency}, tunable electrical and thermal conductivity~\cite{plata2022charting}, and remarkable optoelectronic charecteristics~\cite{rife1977optical}. The presence of heavy metals (Cu, Ag, Be, etc.), coupled with their tunable electronic structure, adds an intriguing dimension from a scientific perspective. The {\it{ab-initio}} calculations for these ternary chalcopyrite semiconductors, characterized by a complex electronic structure, are essential and have been of prime importance.
However, the {\it{ab-initio}} calculations have long predicted that the band gap will be strongly influenced by the anion displacement ($u$) from its primary mean position~\cite{vidal2010strong,siebentritt2010electronic}.

Experimentally,
acquiring the band gaps and optical spectra of Cu and Ag chalcopyrite systems typically involves techniques like photoemission or photoluminescence spectra~\cite{alonso2001optical,Shay1972hybridization,Tell1971electrical,Bellabarba1996optical}.
However, from a theoretical standpoint, achieving an accurate treatment of excited states in these systems requires a comprehensive approach. This often involves either a fully relativistic treatment of many-body perturbation theory (MBPT) or a fully self-consistent Green function-based method ($GW$)~\cite{Hedin1965GW,Aryasetiawan1998GW} accompanied by the solution of the Bethe-Salpeter equation (BSE)~\cite{BSE}. While approaches like fully self-consistent $GW$ and BSE@$GW$ encompass crucial many-body effects, such as electron-electron($e-e$) and electron-hole($e-h$) interactions~\cite{Aryasetiawan1998GW,Schilfgaarde2006quasiparticle,Kotani2007Quasiparticle,Onida2002electronic}, they are known to be computationally demanding and technically challenging, posing significant hurdles to their widespread application.


Importantly, there are several limitations when higher-level methods
are applied to study the electronic structure of chalcopyrite systems.
Firstly, the strong $p-d$ hybridization near the top of the valence band closes the Kohn-Sham (KS) bandgap,  causes a divergence in the dipole transition matrix element, which
poses a significant challenge in constructing a reliable response function in
$GW$ or BSE@$GW$ calculations.
Consequently, these higher level-methods strongly depend on the chosen DFT wavefunctions and orbital energies, and an inaccurate estimation of dielectric function may lead to a wrong exciton effects~\cite{Zhang2013near,Irene2011first}.
Secondly, the band gaps of chalcopyrite systems depend crucially on the
parameters related to the crystal structure,
in particular, on anion displacement ($u$)~\cite{Irene2011first,vidal2010strong}. Thus, the incorrect estimation of the structural parameters
can affect the excitonic wave function.
Consequently, due to these factors, unphysical absorption peaks may appear in the {\it{ab-initio}} optical spectrum.

Nevertheless, there is an emerging alternative method for tackling inherently complex issues related to excited state electronic properties. This novel method is based on the solution of Kohn-Sham (KS) equation or a generalized KS scheme using the dielectric dependent hybrid functional (DDH)~\cite{BrawVorosGovo2016,WeiGiaRigPas2018,BrawGovoVoro2017,ZhengGovoniGalli2019,jana2023simple}.
In this paper, we employ dielectric dependent hybrid functionals to investigate the electronic structure and optical properties of chalcopyrite semiconductors. The DDH functional scheme is proposed as a viable alternative to the highly demanding and complex $GW$ and BSE@$GW$ schemes, offering a more accessible approach for exploring various
systems~\cite{GeroBottCaraOnid2015,GeroBottCara2015,MicelChenIgor2018,ZhengGovoniGalli2019,GeroBottValeOnid2017,WeiGiaRigPas2018,HinuKumaTana2017,BrawGovoVoro2017,LiuCesaMart2020,OhadWingGant2022,WingHabeJonah2019,AshwinDahvydLeeor2019,Wing2019comparing,kronik2018dielectric,Ashwin2019Transferable,jana2023simple,Camarasa2023Transferable}.
Notably, to the best of our knowledge, no study has been reported on the performance of the DDH functional in the case of ternary chalcopyrite semiconductors.

The findings presented in our work strongly suggest the efficacy of DDH functional as a state-of-the-art scheme for accurately characterizing both ground state and excited state electronic properties in highly localized $d$ electronic systems, particularly within the chalcopyrite framework where describing the robust hybridization between $p$ and $d$ orbitals is crucial. {\textcolor{black}{While modern meta-generalized gradient approximations (meta-GGAs) outperform generalized gradient approximations (GGAs) in terms of structural and band gap properties of semiconductors~\cite{jana2018assessing,patra2019relevance,jana2021szs,patra2021efficient,Tran2021bandgap,ghosh2022efficient,Lebeda2023right}}}, they are plagued by well-known issues like many-electron self-interaction and delocalization errors~\cite{Perdew1981self,CoheMoriYang2012,Mori2008localization,jana2018many}, sometimes inadequate in capturing the degree of electron correlation~\cite{ghosh2021improved,patra2021correct,ghosh2022correct}.

This paper evaluates the accuracy of the non-empirical screened DDH functional in predicting band gaps and optical absorption spectra of chalcopyrite semiconductors. Our results indicate that screened-DDH, addressing the generalized Kohn-Sham (KS) scheme, agrees reasonably well with experimental findings for both band gaps and optical spectra. Notably, the functional is particularly effective for challenging Cu-based chalcopyrites. Our calculations demonstrate the applicability of this method in describing the $p-d$ hybridization where other high-level methods are either insufficient or computationally demanding.

\section{Background of Methodologies}
The Coulomb attenuated method (CAM) of the two-electron operator is used to construct the screened range-separated hybrid (SRSH) as~\cite{kronik2018dielectric},
\begin{equation}
\frac{1}{r_{ij}}=\frac{\alpha+\beta~erf(\mu r_{ij})}{r_{ij}}+\frac{1-[\alpha+\beta~erf(\mu r_{ij})]}{r_{ij}}~,
\label{hy-eq1}
\end{equation}
where $r_{ij}=|{\bf{r}}_i-{\bf{r}}_j|$ is the relative position of two electrons. Here, $0\leq \alpha \leq 1$ and $0 \leq\beta \leq 1$ controls the amounts of Fock (non-local) and semilocal ($sl$) GGA (PBE in this case)~\cite{perdewPRL96} that are mixed to the full exchange-correlation (XC) functional in its long-range ($lr$) or short-range ($sr$). $\mu$ is the screening parameter determined later in this paper. Using Eq.~\ref{hy-eq1}, the resultant generalized form of the  SRSH XC expression becomes,
\begin{equation}
\begin{split}
 E_{xc}^{SRSH}(\alpha,\beta;\mu)=(1-\alpha)E_x^{sr-sl,\mu}+\alpha E_x^{Fock-sr,\mu}\\
            +[1-(\alpha+\beta)]E_x^{sl-lr,\mu}+(\alpha+\beta)E_x^{Fock-lr,\mu}+E_c^{sl}
 \label{hy-eq2}
\end{split}
 \end{equation}
 and the corresponding potential,
 \begin{eqnarray}
 V_{xc}^{SRSH}(\alpha,\beta;\mu)
&=&[\alpha+\beta Erf(\mu r)]V_x^{Fock}+\beta V_x^{sl-sr,\mu} \nonumber\\
&+& [1-(\alpha+\beta)]V_x^{sl} + V_c^{sl}~.
 \label{hy-eq3}
 \end{eqnarray}
 Here, $E_c$ and $V_c$ are the PBE correction energy and potential functionals, respectively. Eq.~\ref{hy-eq3} can be seen as the generalized form of the ``CAM'' type hybrid used extensively for finite and extended systems~\cite{GeroBottCaraOnid2015,GeroBottCara2015,MicelChenIgor2018,ZhengGovoniGalli2019,GeroBottValeOnid2017,WeiGiaRigPas2018,HinuKumaTana2017,BrawGovoVoro2017,LiuCesaMart2020,OhadWingGant2022,WingHabeJonah2019,AshwinDahvydLeeor2019,Wing2019comparing,kronik2018dielectric,Ashwin2019Transferable,jana2023simple,Camarasa2023Transferable}. However, the naming of the functionals becomes different based on how one determines the parameters. Typically, one can consider  $\alpha+\beta=\gamma$, where $\gamma$ is an another parameter. In terms of $\alpha$ and $\gamma$ Eq.~\ref{hy-eq3} becomes,
 \begin{eqnarray}
 V_{xc}^{SRSH}(\alpha,\gamma;\mu)&=&[\alpha-(\alpha-\gamma)Erf(\mu r)]V_x^{Fock}\nonumber\\
 &-&(\alpha-\gamma)V_x^{sl-sr,\mu}+(1-\gamma)V_x^{sl}+ V_c^{sl}~.
 \label{hy-eq-gen}
 \end{eqnarray}

In particular, the following choices are important for bulk solids: (i) for $\alpha=0.25$, $\gamma=0$, and $\mu=0.11$ Bohr$^{-1}$ in Eq.~\ref{hy-eq-gen}, resulting in the recovery of the HSE06 like functionals~\cite{heyd2003hybrid,krukau2006influence,heyd2004efficient,jana2020screened,jana2020improved,jana2018efficient,jana2018meta,jana2019screened},
 \begin{eqnarray}
 V_{xc}^{HSE06}(0.25,0;0.11)&=&0.25[1-Erf(0.11 r)]V_x^{Fock}\nonumber\\
 &-&0.25V_x^{sl-sr,0.11}+V_x^{sl}+ V_c^{sl}~,\nonumber\\
 \label{hy-eq-hse}
 \end{eqnarray}
 and (ii) for $\alpha=1$ and  $\gamma=\epsilon_{\infty}^{-1}$, where $\epsilon_{\infty}$ is the high-frequency macroscopic static dielectric constant or ion-clamped static (optical) dielectric constant or electronic dielectric constant, the resultant functional becomes,
\begin{eqnarray}
 V_{xc}^{DDH}(1,\epsilon_{\infty}^{-1};\mu)&=&[1-(1-\epsilon_{\infty}^{-1})Erf(\mu r)]V_x^{Fock}\nonumber\\
 &-&(1-\epsilon_{\infty}^{-1})V_x^{sl-sr,\mu}+(1-\epsilon_{\infty}^{-1})V_x^{sl}+ V_c^{sl}~,\nonumber\\
 \label{hy-eq6}
 \end{eqnarray}
which is named dielectric-dependent range-separated hybrid functional based
on the CAM (DD-RSH-CAM)~\cite{WeiGiaRigPas2018} or simply DDH (used throughout this paper).

In particular, the model dielectric function for bulk solid is defined according to Eq.~\ref{hy-eq-gen} with $\gamma=\epsilon_\infty^{-1}$ as,
\begin{equation}
 \varepsilon^{-1}(|{\bf{G}}|)=\alpha-(\alpha-\epsilon^{-1}_{\infty})e^{-|{\bf{G}}|{^2}/(4\mu)}~,
 \label{eqsec1-6}
\end{equation}
which is the key to the DDH functional (where ${\bf{G}}$ is the reciprocal lattice
vector). The model dielectric function
makes this construction quite similar to that of the self-energy correction of $GW$, in particular when GGA approximates $\sum_{COH}$ (Coulomb hole (COH)) and $\sum_{SEX}$ (screened exchange (SEX)) by Fock term~\cite{CuiWangZhang2018}.

It is readily apparent from Eq.~\ref{hy-eq6} that the macroscopic static dielectric constant, $\epsilon_\infty$, is the key to DDH calculations. It can be obtained using different procedures discussed in section~\ref{sec-dielectric}. Several procedures are also available to determine the screening parameter $\mu$. In particular, $\mu$ can be obtained (i) depending on the valance electron density that
participate to the screening~\cite{BrawVorosGovo2016,SkonGovoGall2016,CuiWangZhang2018},
(ii) from fitting with the accurate dielectric function~\cite{WeiGiaRigPas2018}, (iii) from empirical fitting~\cite{Yang2023range} or (iv) from first principle way using linear-response TDDFT (LR-TDDFT) approaches based on local density or local Seitz radius ($r_s$)~\cite{jana2023simple}. In particular, in this paper, we use procedure (iv) to determine $\mu$, which is named as $\mu_{eff}^{fit}$ and obtained using the compressibility sum rule together with LR-TDDFT~\cite{jana2023simple} having the form,
\begin{equation}
 \mu=\mu_{eff}^{fit}= \frac{a_1}{\langle r_s \rangle} + \frac{a_2 \langle r_s \rangle}{1 + a_3 \langle r_s \rangle^2}~,
 \label{eq-theo-secb-19}
\end{equation}
with $a_1= 1.91718$, $a_2= -0.02817$, $a_3=0.14954$, and
\begin{equation}
 \langle r_s \rangle=\frac{1}{V_{cell}}\int_{cell} \Big(\frac{3}{4\pi(n_{\uparrow}({\bf{r'}})+n_{\downarrow}({\bf{r'}}))}\Big)^{1/3}~d^3r'~.
 \label{eq-theo-secb-13}
\end{equation}
The readers are referred to ref.~\cite{jana2023simple} for the details of this formula and underlying derivations. It may be noted
that the resultant
$\mu_{eff}^{fit}$ performs quite similarly
to those obtained from the fitting of the dielectric function
~\cite{WeiGiaRigPas2018}
as reported in ref.~\cite{jana2023simple}.
On the other hand, the screening parameters determined from procedure (i) are not always well-defined in some materials, especially where electrons of different characters participate in the
valence bands~\cite{Lorin2021first}. In this respect, determining screening parameters from method (iv) is quite reasonable and well justified~\cite{jana2023simple}.



{\textcolor{black}{In the subsequent discussion, we briefly overview the different levels within the $GW$, which we have used whenever applicable. Various approximations exist for $GW$, including the single-shot $G_0W_0$ calculation, which heavily relies on the choice of the initial KS wavefunctions and orbital energies~\cite{Gant2022optimally}. On the other hand, $GW_0$ involves a self-consistent update of the orbital energies in the Green's function $G$ after the initial $G_0W_0$ step~\cite{Hybertsen1986electron,Shishkin2007accurate}.
One may note the crucial distinctions between $GW$ and DDH. The $GW$ steps involve the calculations of the frequency dependent dielectric function, including the summations over both occupied and unoccupied states. This makes the self-consistent of $GW$ computationally more expensive.}} In contrast, the DDH calculations are performed only within the generalized KS (gKS) scheme. However, it requires an additional calculation of dielectric constants (details are provided in sec.~\ref{secIII-C}). Second, the outcomes of one-shot $G_0W_0$ and partially self-consistent $GW_0$ are highly dependent on the initial choice of the KS functional, while DDH, being a self-consistent approach within gKS, yields outcomes independent of the initial state.




\section{Results and Discussions}

\subsection{Materials and calculation details}
\label{mat-cal}

\subsubsection{Materials}
\label{mat}

\begin{table}
\begin{center}
\caption{\label{tab-all-0} PBE optimized $a$ (in \AA), $c$, $\eta$, $r_{A-X}$ (in \AA), $r_{B-X}$ (in \AA), and $u$ of all the chalcopyrite semiconductors used in this work.}
\begin{ruledtabular}
\begin{tabular}{lcccccccccccccccccccccccc}
		Solids	& $a$ & c & $\eta$	& $r_{A-X}$ & $r_{B-X}$ & $u$ \\
 \hline\\
 \multicolumn{7}{c}{I-III-VI$_2$}\\[0.2 cm]
        AgAlS$_2$ & 5.740 & 5.251 & 0.915 & 2.574 & 2.277 & 0.294 \\
        AgAlSe$_2$ & 6.029 & 5.562 & 0.922 & 2.677 & 2.425 & 0.285 \\
        AgAlTe$_2$ & 6.409 & 6.122 & 0.955 & 2.812 & 2.661 & 0.270 \\
        AgGaS$_2$ & 5.773 & 5.307 & 0.919 & 2.563 & 2.317 & 0.286 \\
        AgGaSe$_2$ & 6.049 & 5.632 & 0.931 & 2.665 & 2.463 & 0.278 \\
        AgGaTe$_2$ & 6.406 & 6.169 & 0.963 & 2.801 & 2.681 & 0.266 \\
        AgInS$_2$ & 5.925 & 5.749 & 0.970 & 2.573 & 2.509 & 0.259 \\
        AgInSe$_2$ & 6.195 & 6.039 & 0.975 & 2.673 & 2.647 & 0.254 \\
        AgInTe$_2$ & 6.567 & 6.500 & 0.990 & 2.812 & 2.856 & 0.244 \\
        CuAlS$_2$ & 5.336 & 5.274 & 0.988 & 2.324 & 2.28 & 0.257 \\
        CuAlSe$_2$ & 5.651 & 5.576 & 0.987 & 2.443 & 2.430 & 0.252 \\
        CuAlTe$_2$ & 6.094 & 6.055 & 0.994 & 2.603 & 2.664 & 0.241 \\
        CuGaS$_2$ & 5.372 & 5.315 & 0.989 & 2.314 & 2.322 & 0.249 \\
        CuGaSe$_2$ & 5.677 & 5.631 & 0.992 & 2.432 & 2.471 & 0.244 \\
        CuGaTe$_2$ & 6.096 & 5.086 & 0.834 & 2.593 & 2.685 & 0.237 \\
        CuInS$_2$ & 5.578 & 5.617 & 1.007 & 2.33 & 2.519 & 0.220 \\
        CuInSe$_2$ & 5.871 & 5.908 & 1.006 & 2.447 & 2.657 & 0.219 \\
        CuInTe$_2$ & 6.294 & 6.317 & 1.003 & 2.608 & 2.861 & 0.215 \\[0.2 cm]
        \multicolumn{7}{c}{II-IV-V$_2$}\\[0.2 cm]
        BeGeAs$_2$ & 5.446 & 5.48 & 1.006 & 2.267 & 2.468 & 0.218 \\
        BeGeP$_2$ & 5.207 & 5.229 & 1.004 & 2.177 & 2.347 & 0.222 \\
        BeSiAs$_2$ & 5.372 & 5.368 & 0.999 & 2.266 & 2.388 & 0.230 \\
        BeSiP$_2$ & 5.129 & 5.117 & 0.998 & 2.166 & 2.266 & 0.233 \\
        BeSnAs$_2$ & 5.655 & 5.697 & 1.007 & 2.287 & 2.650 & 0.194 \\
        BeSnP$_2$ & 5.428 & 6.675 & 1.230 & 2.198 & 2.538 & 0.195 \\
        CdGeAs$_2$ & 6.052 & 5.746 & 0.949 & 2.676 & 2.485 & 0.277 \\
        CdGeP$_2$ & 5.805 & 5.484 & 0.945 & 2.583 & 2.363 & 0.282 \\
        CdSiAs$_2$ & 5.979 & 5.551 & 0.928 & 2.676 & 2.396 & 0.289 \\
        CdSiP$_2$ & 5.728 & 5.289 & 0.923 & 2.584 & 2.271 & 0.296 \\
        CdSnAs$_2$ & 6.218 & 6.103 & 0.981 & 2.688 & 2.664 & 0.253 \\
        CdSnP$_2$ & 5.983 & 5.862 & 0.980 & 2.597 & 2.550 & 0.257 \\
        MgGeAs$_2$ & 6.020 & 5.638 & 0.936 & 2.637 & 2.473 & 0.273 \\
        MgGeP$_2$ & 5.796 & 5.365 & 0.926 & 2.551 & 2.355 & 0.279 \\
        MgSiAs$_2$ & 5.968 & 5.410 & 0.906 & 2.638 & 2.385 & 0.286 \\
        MgSiP$_2$ & 5.746 & 5.125 & 0.892 & 2.555 & 2.265 & 0.292 \\
        MgSnAs$_2$ & 6.152 & 6.042 & 0.982 & 2.642 & 2.654 & 0.248 \\
        MgSnP$_2$ & 5.935 & 5.799 & 0.977 & 2.558 & 2.543 & 0.252 \\
        ZnGeP$_2$ & 5.500 & 5.420 & 0.985 & 2.383 & 2.357 & 0.254 \\
        ZnSiAs$_2$ & 5.676 & 5.530 & 0.974 & 2.480 & 2.395 & 0.263 \\
        ZnSiP$_2$ & 5.420 & 5.268 & 0.972 & 2.383 & 2.270 & 0.268 \\
        ZnSnAs$_2$ & 5.944 & 5.961 & 1.003 & 2.498 & 2.659 & 0.226 \\
        ZnSiP$_2$ & 5.703 & 5.718 & 1.003 & 2.402 & 2.545 & 0.228 \\
    \end{tabular}
\end{ruledtabular}
\end{center}
\end{table}
%

We use $42$ chalcopyrite semiconductors having ABX$_2$ structures (with space group $I{\bar{42}}d$) grouped as I-III-VI$_2$ (18 chalcopyrites) and I-III-VI$_2$ (24 chalcopyrites), which are the iso-electronic analogous of the II$-$VI and III$-$V ideal zinc blende structure (distorted), respectively~\cite{shaposhnikov2012abinitio}. Here, $(A, B)$ are the two
cations tetrahedrally coordinated by four anions ($X$), where each anion is again coordinated by two cations each of which are $A$ and $B$ types~\cite{jaffe1983electronic}. Three structural parameters, namely, ($i$) lattice constant $a$, ($ii$) tetragonal ratio $\eta=c/2a$
where $c$  is the lattice constant along the $z$-direction, and ($iii$) the anion displacement parameter, $u$ are used to describe a chalcopyrite structure. Note that $u$ is an important structural parameter used to describe physics related to the interplay between structure and electronic properties and is defined as~\cite{vidal2010strong},
\begin{equation}
u=0.25+(r^2_{A-X}-r^2_{B-X})/a^2~,
\label{eqq1}
\end{equation}
where $r_{A-X}$ and $r_{B-X}$ are $A-X$ and $B-X$ bond lengths, respectively.

DDH, HSE06, and $GW_0$ band gap calculations are performed using the PBE optimized geometries. The details of the PBE optimized geometries are given in Table~\ref{tab-all-0}. In general, the computed lattice constants from
PBE XC functionals agree well with the experimental.

\subsubsection{Calculation details}

The density functional calculations are performed using the plane-wave formalism as implemented in {\it{Vienna Ab initio Simulation Package}} (VASP) code~\cite{vasp1,vasp2,vasp3,vasp4}.
A kinetic energy cutoff of $520$ eV is used for all DFT calculations. We use  Monkhorst-Pack (MP) like $\Gamma-$centered $11\times11\times11$ {\bf{k}}-points mesh to sample the Brillouin Zone (BZ) for PBE calculations, whereas for DDH and HSE06, the {\bf{k}}-points are reduced to $8\times8\times8$. The electronic energies are allowed to converge at $10^{-6}$ eV for all DFT methods to achieve self-consistency. The relaxation of the structures is performed till the
Hellmann-Feynman forces on atoms are reduced to less than 0.01 eV/\AA$^{-1}$. The VASP-recommended PAW pseudopotentials are used. Noteworthy, relatively deep Ga $3d$, Ge $3d$,
and In $4d$, states are treated as valence orbitals.

We also perform partially self-consistent quasi-particle $GW_0$ calculations from the VASP code whenever required for bandgaps. The implementation of this method in the VASP is described in ref.~\cite{Shishkin2007accurate}. For the QP calculations, the number of virtual orbitals is increased to $240$ (using NBANDS=$240$), and $4$ iterations for the self-consistent $GW_0$ steps are used (using NELMGW=$4$) after $G_0W_0$ calculation. The VASP-recommended $GW$ pseudopotentials are used, where relatively deep Ga $3d$, Ge $3d$,
and In $4d$ states are treated as valence orbitals. $8\times8\times8$ $\Gamma-$centered $k-$ points are used to sample the Brillouin zone in $GW_0$ calculations. For all our cases, the starting point of many-body perturbation theory is the GGA PBE functional.

The optical absorption spectrum  for DDH and HSE06 is also performed using the VASP code with a
$16 \times 16 \times 16$ MP-like $\Gamma-$centered ${\bf{k}}$-points with $72$ empty orbitals. We have performed the DDH and HSE06 calculations in many
shifted $4 \times 4 \times 4$ ${\bf{k}}$-points and weight over the multiple grids as a straightforward calculation of hybrids would be expensive.~\cite{vaspImprovingDielectric}. 

\begin{figure}
\begin{center}
\includegraphics[width=\columnwidth]{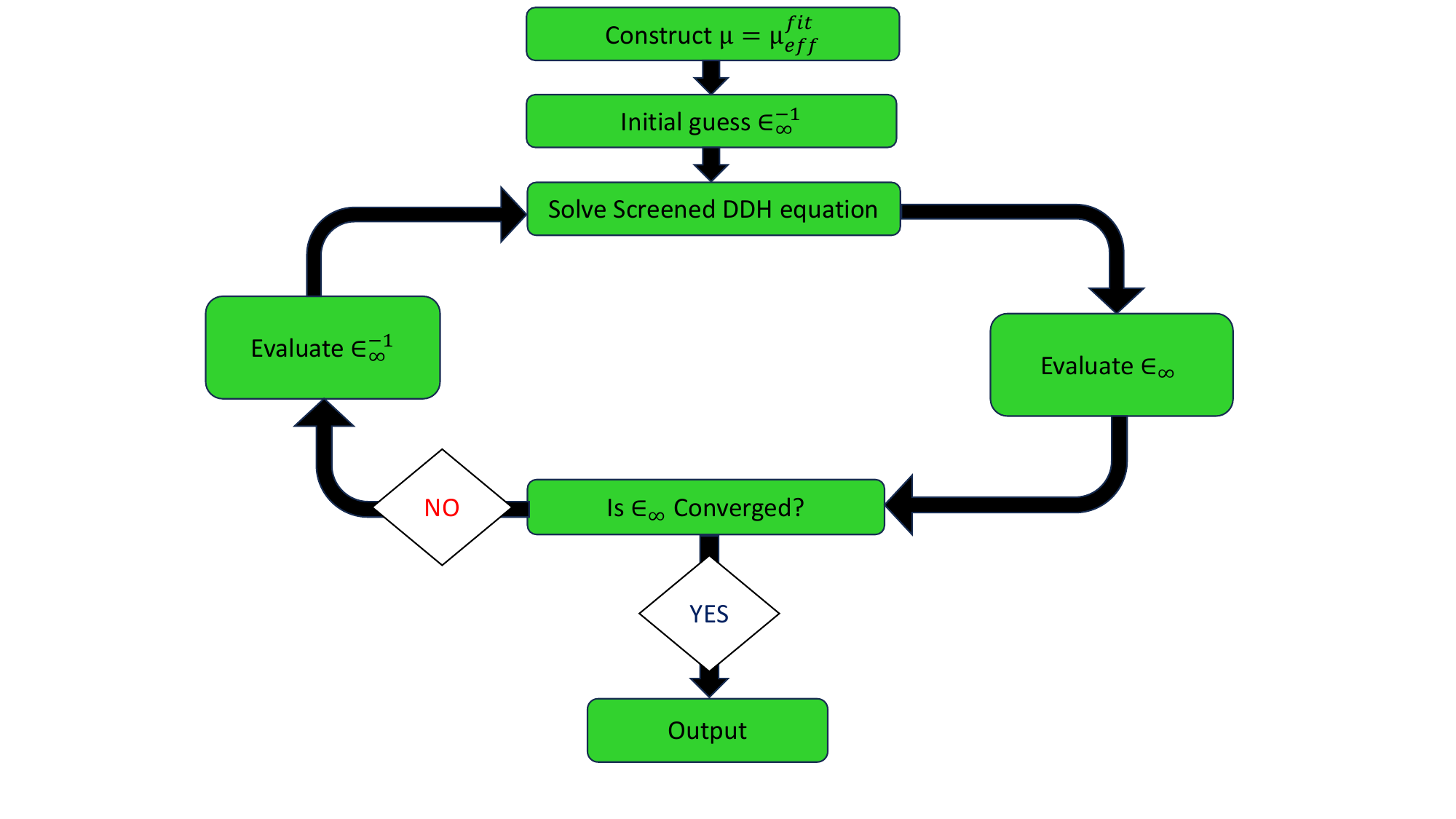}
\end{center}
\caption{\label{fig-ddh-scf}  Pictorial representation of the self-consistency of the screened DDH functional used in this paper.}
\end{figure}
\begin{figure}
\begin{center}
\includegraphics[width=\columnwidth]{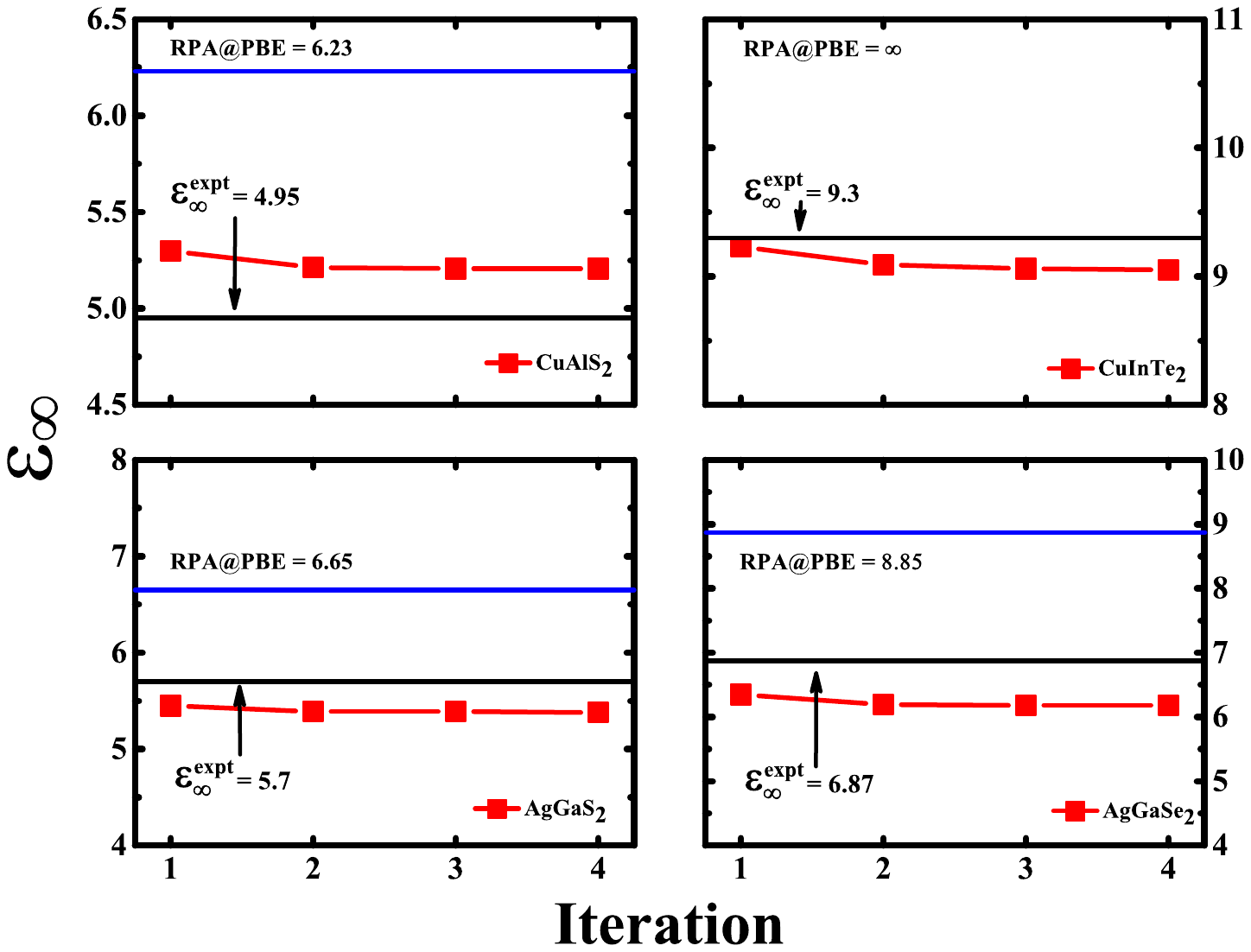}
\end{center}
\caption{\label{fig-iteration}  Convergence of $\epsilon_\infty$ as obtained using RPA@DDH using the self-consistent cycle of Fig.~\ref{fig-ddh-scf}. See Fig.~\ref{fig-epsilon-comp-expt} for the experimental values.}
\end{figure}
\begin{figure}
\begin{center}
\includegraphics[width=\columnwidth]{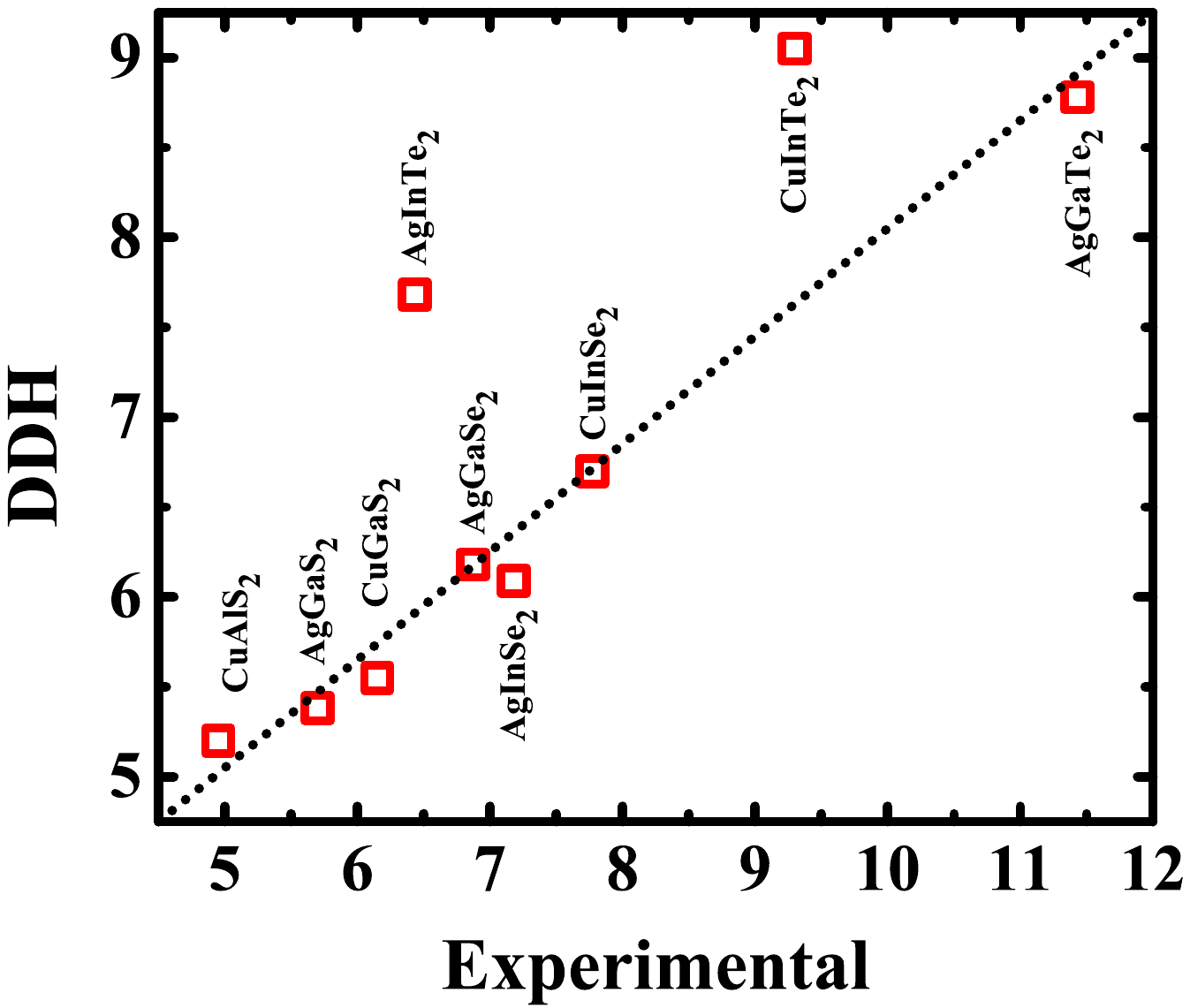}
\end{center}
\caption{\label{fig-epsilon-comp-expt}   The orientationally averaged static dielectric constants (optical), $\epsilon_\infty$ from experimental and RPA@DDH for nine Cu and Ag-based chalcopyrite. The self-consistent cycle of Fig.~\ref{fig-ddh-scf} is used for RPA@DDH. The experimental values are used 7.77 for CuInSe$_2$~\cite{Neumann1989experimetal}, 9.30 for CuInTe$_2$\cite{Holah1981infrared}, 6.15 for CuGaS$_2$~\cite{Baars1972dielectric}, 4.95 for CuAlS$_2$~\cite{Koschel1973solid}, 7.18 for AgInSe$_2$~\cite{Kanell1977}, 6.43 for AgInTe$_2$~\cite{Kanell1977}, 6.87 for AgGaSe$_2$~\cite{Madelung1992}, 5.70 for AgGaS$_2$~\cite{Ziel1974lattice}, 11.43 for AgGaTe$_2$~\cite{Madelung1992}.
}
\end{figure}
\begin{figure}
\begin{center}
\includegraphics[width=\columnwidth]{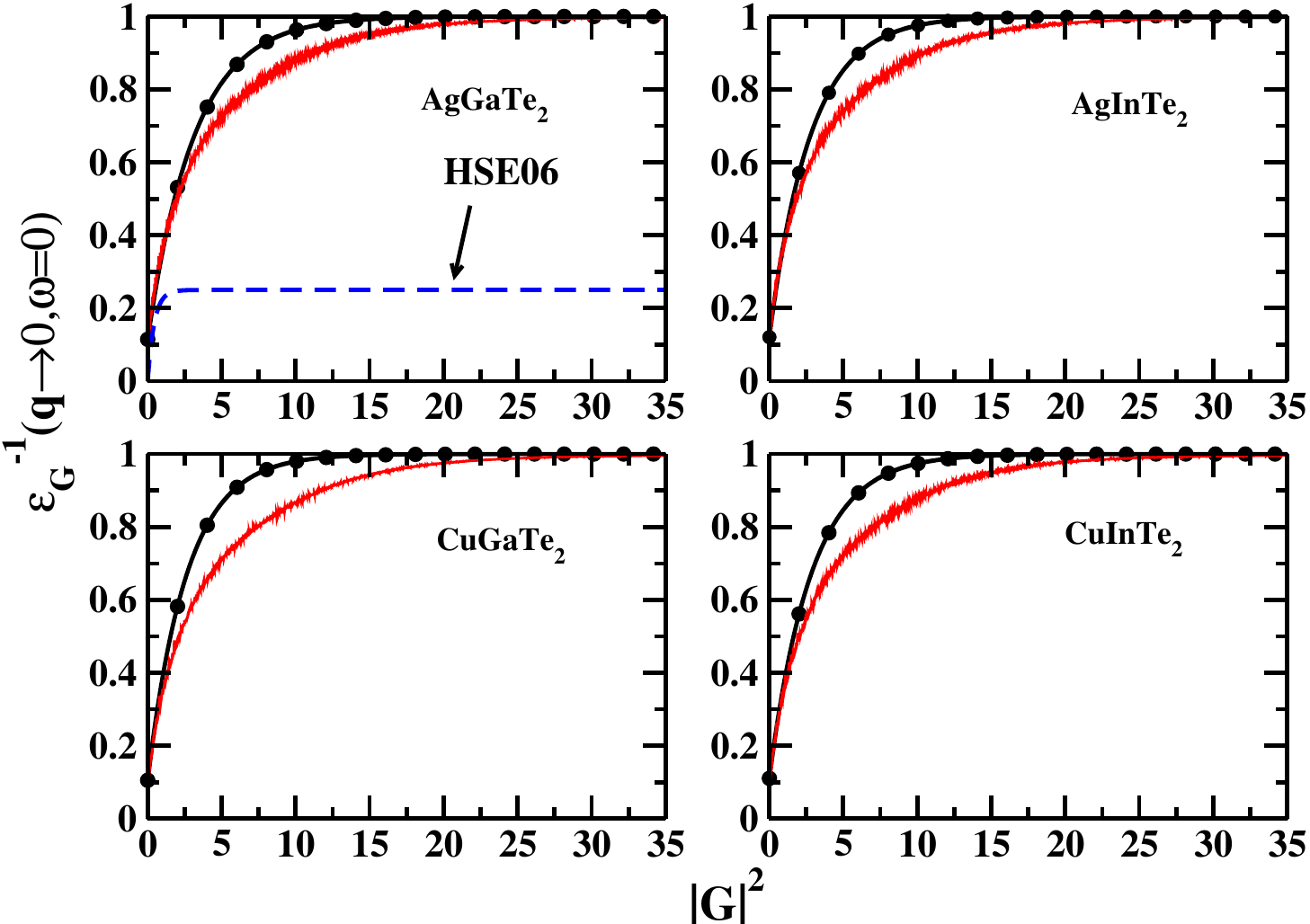}
\end{center}
\caption{\label{die-comp}  Comparison of model dielectric function as a function of ${\bf{G}}$ (in Bohr$^{-1}$) for four chalcopyrite semiconductors. The black circle line indicates the model function with parameters evaluated at RPA@DDH (see Table~\ref{tab-bg}), whereas the red line corresponds to the $G_0W_0$ or RPA calculations.}
\end{figure}

\subsection{Self-consistent screened-DDH calculation}
\label{sec-dielectric}

The central quantity of the screened DDH is to determine high-frequency macroscopic static dielectric constants or ion-clamped static (optical) dielectric constant or electronic dielectric constant, $\epsilon_\infty$ and screening parameter, $\mu$. As mentioned before, in our case, we use $\mu=\mu_{eff}^{fit}$ of Eq.~\ref{eq-theo-secb-19}. The calculations of $\mu_{eff}^{fit}$ are done using the LDA densities, and those are used further for all the chalcopyrite systems. These values are supplied in Table~\ref{tab-bg}. As shown in Table~\ref{tab-bg}, $\mu_{eff}^{fit}$'s remain almost constants to a typical average value $0.72$ bohr$^{-1}$,  although for a few systems, its value becomes $\sim$ $0.80$ Bohr$^{-1}$. These values are consistent with the earlier reported 
values for common semiconductors and insulators~\cite{WeiGiaRigPas2018,jana2023simple}. Noteworthy, in ref.~\cite{WeiGiaRigPas2018}, $\mu$'s are calculated from the least-squares fitting to the dielectric function in the long-wavelength limit obtained from higher-level accurate calculations, such as random phase approximation (RPA) calculated using PBE orbitals (RPA@PBE) or using nanoquanta kernel and partially self-consistent GW calculations~\cite{Shishkin2007accurate}.

Next, we turn to the calculations of the static dielectric constant, which is the central quantity for any DDH. As for our present case, we already fixed the $\mu$ to $\mu_{eff}^{fit}$ from the scheme described above. Considering $\epsilon_\infty$, it can be calculated using various schemes: (i) density functional perturbation theory (DFPT) within the framework of the RPA~\cite{Baroni2001phonons} or (ii) using the modern theory of polarization~\cite{Smith1993theory,Resta1994macroscopic}.

The fundamental equation that has been solved to calculate the dielectric tensor for bulk solid is,
\begin{equation}
\epsilon_{{\bf{G}},{\bf{G'}}}({\bf{q}},\omega)=\delta_{{\bf{G}},{\bf{G'}}}-\frac{4\pi}{|{\bf{G+q}}||{\bf{G'+q}}|}\chi_{{\bf{G}},{\bf{G'}}}({\bf{q}},\omega)~.
\label{eq-epsilon-rpa}
\end{equation}
Our interest lies on the inverse of the macroscopic dielectric matrix $\epsilon^{-1}_\infty({\bf{q}},\omega)$, which is obtained from the head of the inversion of the full microscopic dielectric tensor as,

\begin{equation}
\epsilon^{-1}_\infty({\bf{q}},\omega)=\lim_{{\bf{q}}\to 0}\epsilon^{-1}_{0,0}({\bf{q}},\omega)
\end{equation}
In Eq.~\ref{eq-epsilon-rpa}, the reducible polarizability $\chi_{{\bf{G}},{\bf{G'}}}({\bf{q}},\omega)$ is defined as,
\begin{equation}
 \chi_{{\bf{G}},{\bf{G'}}}=\chi^0_{{\bf{G}},{\bf{G'}}}+\chi^0_{{\bf{G}},{\bf{G'}}}\Big(\frac{4\pi}{|{\bf{G+q}}||{\bf{G'+q}}|}+f_{xc}({\bf{q}},\omega)\Big)\chi_{{\bf{G}},{\bf{G'}}}~,
\end{equation}
where 
$\chi^0_{{\bf{G}},{\bf{G'}}}({\bf{q}},\omega)$ being the irreducible polarizability matrix obtained from KS systems~\cite{Gajdo2006linear,Nunes2001berry,Souza2002first} and $f_{xc}({\bf{q}},\omega)$ is the XC kernel, obtained from the derivative of the exchange-correlation potential~\cite{Baroni2001phonons}. However, $f_{xc}({\bf{q}},\omega)$ can be neglected as its inclusion in the polarizability calculation is observed to be negligible~\cite{paier2008dielectric}. Henceforth, in this work, the $\chi_{{\bf{G}},{\bf{G'}}}$ and dielectric constants are evaluated by neglecting $f_{xc}({\bf{q}},\omega)$, i.e., within RPA approximations. Noteworthy, using Bloch notation, one can write $\chi^0_{{\bf{G}},{\bf{G'}}}({\bf{q}},\omega)$ in terms of the KS orbitals and energies~\cite{Adler1962quantum,wiser1963dielectric,Gajdo2006linear}. Therefore, the evaluation of $\epsilon_\infty({\bf{q}},\omega)$ depends strongly on the choices of the XC functionals. Here, we perform RPA calculations of Eq.~\ref{eq-epsilon-rpa} with PBE orbitals (named RPA@PBE) and DDH orbitals (named RPA@DDH).

Finally, the self-consistent cycle to calculate the static $\epsilon_\infty({\bf{q}}\to 0,\omega\to 0)$ using screened-DDH is described in Fig.~\ref{fig-ddh-scf}. We conform to the following steps: (i) Firstly, we calculate $\mu_{eff}^{fit}$ using the Eq.~\ref{eq-theo-secb-19} with LDA orbitals, (ii) secondly, we start with $\epsilon_\infty({\bf{q}}\to 0,\omega\to 0)$ as obtained from PBE functional (RPA@PBE), and plug it in our DDH expression Eq.~\ref{hy-eq6} along with previously calculated $\mu_{eff}^{fit}$, (iii) thirdly, we
perform the DDH calculation and update $\epsilon_\infty({\bf{q}}\to 0,\omega\to 0)$ as obtained from RPA@DDH as long as the self-consistency in $\epsilon_\infty({\bf{q}}\to 0,\omega\to 0)$ is reached. In Fig.~\ref{fig-iteration} we illustrate the self-consistency of $\epsilon_\infty({\bf{q}}\to 0,\omega\to 0)$ using RPA@DDH using the scheme of Fig.~\ref{fig-ddh-scf}.
The self-consistency of $\epsilon_\infty({\bf{q}}\to 0,\omega\to 0)$ is achieved mostly within four cycles.
It's worth mentioning that for certain chalcopyrites, the PBE predicts a metallic nature with zero bandgaps, leading to $\epsilon_\infty({\bf{q}}\to 0, \omega\to 0)\to \infty$ at RPA@PBE. In such cases, one needs to initiate the self-consistency process illustrated in Fig.~\ref{fig-ddh-scf} from a finite value of $\epsilon_\infty({\bf{q}}\to 0,\omega\to 0)$. Noteworthy, the potential is solved in all materials using the generalized Kohn-Sham (gKS) scheme~\cite{Garrick2020exact}.

\subsection{High-frequency dielectric constants}
\label{secIII-C}
In Table~\ref{tab-bg}, we first calculate the orientationally-averaged $\epsilon_\infty$ (i.e., $\epsilon_\infty=\frac{\epsilon^{xx}_\infty+\epsilon^{yy}_\infty+\epsilon^{zz}_\infty}{3}$) with RPA using PBE and DDH XC approximations with $\mu$ using Eq.~\ref{eq-theo-secb-19}.
%
%
The analysis from Table~\ref{tab-bg} reveals that, for I-III-VI$2$, RPA@PBE tends to overestimate $\epsilon\infty$ compared to RPA@DDH. Notably, for several cases like AgGaTe$_2$, AgInTe$_2$, CuGaSe$_2$, CuInS$_2$, CuGaTe$_2$, and CuInTe$2$, RPA@PBE predicts significantly large $\epsilon\infty$, particularly in instances where PBE incorrectly predicts a metallic structure. The dielectric constants calculated with RPA@PBE show inaccuracies for such materials.
A similar trend is observed for II-IV-V$_2$ chalcopyrite semiconductors. However, in cases like ZnGeAs$_2$ and ZnSnAs$_2$, PBE calculations lead to a metallic outcome. Across all instances, there is a noticeable magnitude difference of approximately $\sim 2$ when comparing RPA@DDH and RPA@PBE.
Examining the $\epsilon_\infty$ values from RPA@DDH, we observe that all values are finite and fall within the range expected for an ideal chalcopyrite semiconductor. This performance highlights the inadequacy of RPA@PBE in accurately calculating $\epsilon_\infty$ for these systems.



 Next, Fig.\ref{fig-epsilon-comp-expt} provides a comparison of $\epsilon_\infty$ for eleven Cu and Ag-based chalcopyrites obtained from RPA@DDH and experimental data derived from optical reflectivity experiments (refer to the references in the caption of Fig.\ref{fig-epsilon-comp-expt}). We observe a remarkable agreement between the calculated $\epsilon_\infty^{RPA@DDH}$ and the experimental results. Conversely, for most of these systems, PBE-predicted $\epsilon_\infty$ values are quite large, with most systems exhibiting metallic behavior (as shown in Table~\ref{tab-bg}).

Finally, Fig.\ref{die-comp}, we compare the model dielectric function described by Eq.\ref{eqsec1-6} (with $\alpha=1$, $\mu=\mu_{eff}^{fit}$, and self-consistent $\epsilon_\infty^{-1}$ evaluated at RPA@DDH, as per Table~\ref{tab-bg}) with the dielectric function obtained through $G_0W_0$ or RPA calculations for various chalcopyrite semiconductors. The details of the calculation procedure can be found in ref.\cite{vaspImprovingDielectric}. This comparison illustrates that the model dielectric function given by Eq.\ref{eqsec1-6} aligns well with dielectric functions calculated through {\it{ab-initio}} methods.


\begingroup
\begin{table*}
\caption{\label{tab-bg} High-frequency macroscopic static dielectric constants or ion-clamped static (optical) dielectric constant or electronic dielectric constant ($\epsilon_\infty$), screening parameters ($\mu$ in bohr $^{-1}$), and direct KS bandgaps (in eV) using PBE, DDH, and HSE06. Here RPA@DDH is evaluated using the scheme given in Fig.~\ref{fig-ddh-scf}. The PBE, DDH, and HSE06 band gaps are corrected for spin-orbit coupling (SOC). Error statistics in bandgaps with respect to experimental values are also given. The experimental reference values of I-III-VI$_2$ and II-IV-V$_2$ are taken from ref.~\cite{Xiao2011accurate} and ref.~\cite{shaposhnikov2012abinitio}, respectively. Here, the `$-$' lines indicate that those systems are treated as metals when using PBE. The bandgap values close to the experiment one are in bold font. Total ME (in eV), MAE (in eV), and MARE are also shown at the end of the Table.
}
\begin{tabular}{cccccccccccccccccccccccccccccccccc}
\hline\hline
	Solids&&$\epsilon_\infty$ (RPA@PBE)&$\epsilon_\infty$ (RPA@DDH)&&$\mu=\mu_{eff}^{fit}$ &&	E$_g$ (PBE)&E$_g$ (DDH)&E$_g$ (HSE06)&E$_g$ ($GW_0$)&E$_g$ (Expt.)&$\Delta_{SOC}^a$ \\
	\hline\\
\multicolumn{12}{c}{I-III-VI$_2$}\\[0.4 cm] 
AgAlS$_2$	&	&	5.46	&	4.97	&	&	0.78	&	&	1.86	&	{\bf{3.70}}	& 3.05 & 3.32$^b$		&	3.60&0.00	\\
AgAlSe$_2$	&	&	6.64	&	5.70	&	&	0.78	&	&	1.07	&	2.67	& 2.18 &{\bf{2.43}}$^b$		&	2.55&0.03	\\
AgAlTe$_2$	&	&	8.30	&	7.30	&	&	0.73	&	&	0.86	&	1.88	& 1.98 &{\bf{2.16}}$^b$		&	2.30&0.17	\\
AgGaS$_2$	&	&	6.65	&	5.38	&	&	0.82	&	&	1.11	&	{\bf{2.86}}	& 2.30 &2.16$^b$		&	2.73&0.01	\\
AgGaSe$_2$	&	&	8.85	&	6.18	&	&	0.82	&	&	0.51	&	{\bf{2.06}}	& 1.62 &1.29$^b$		&	1.83&0.03	\\
AgGaTe$_2$	&	&	15.39	&	8.78	&	&	0.79	&	&	0.13	&	1.19	& {\bf{1.23}}&1.17$^b$		&	1.36&0.17	\\
AgInS$_2$	&	&	7.56	&	5.21	&	&	0.74	&	&	0.49	&	2.23	& {\bf{1.57}}&1.32$^b$		&	1.87&0.01	\\
AgInSe$_2$	&	&	$-$	&	6.09	&	&	0.74	&	&	0.02	&	{\bf{1.56}}	& 1.06&0.73$^b$		&	1.24&0.04	\\
AgInTe$_2$	&	&	11.8	&	7.68	&	&	0.70	&	&	0.18	&	1.28	& 1.34 &{\bf{0.81}}$^b$		&	1.04& 0.19	\\
CuAlS$_2$	&	&	6.23	&	5.20	&	&	0.78	&	&	1.66	&	3.87	& {\bf{3.20}} &2.96$^b$		&	3.46&0.01	\\
CuAlSe$_2$	&	&	7.73	&	6.05	&	&	0.78	&	&	0.84	&	{\bf{2.76}}	& 2.32 &2.13$^b$ 		&	2.65&0.02	\\
CuAlTe$_2$	&	&   9.42    &	7.85	&	&	0.75	&	&	0.93	&	1.96	& {\bf{2.01}} &1.85$^b$		&	2.06&0.09	\\
CuGaS$_2$	&	&	7.80	&	5.55	&	&	0.70	&	&	0.89	&	3.17	& 2.22 & 1.68$^d$, 1.78$^b$, {\bf{2.35}}$^e$		&	2.50&0.01	\\
CuGaSe$_2$	&	&	13.57	&	6.42	&	&	0.70	&	&	0.27	&	2.33	& 1.55 &0.93$^d$, 0.99$^b$, {\bf{1.60}}$^e$ 		&	1.67&0.04	\\
CuGaTe$_2$	&	&12.54	&	8.54	&	&	0.66	&	&	0.36	&	{\bf{1.45}}	& 1.65 &0.90$^b$		&	1.25&0.18	\\
CuInS$_2$	&	&	$-$	&	5.98	&	&	0.73	&	&	0.03	&	1.72	& 1.03 &0.77$^d$, 0.69$^b$, {\bf{1.41}}$^e$ 		&	1.55&-0.03	\\
CuInSe$_2$	&	&	17.41	&	6.70	&	&	0.73	&	&	0.01	&	1.48	& 0.88 & 0.46$^b$, {\bf{0.93}}$^e$ 		&	1.04&0
.01\\
CuInTe$_2$	&	&	$-$	&	9.05	&	&	0.71	&	&	0.01	&	{\bf{1.14}}	& 1.17 &0.70$^b$		&	1.00& 0.17	\\
	&		&		&		&		&		&		&		\\
\multicolumn{12}{c}{II-IV-V$_2$} \\ [0.4 cm]
BeGeAs$_2$	&	&	11.30   &	10.47	&	&	0.81	&	&	0.55	&	{\bf{1.31}}	& 1.30 &	1.07$^c$	&	1.68&0.07	\\
BeGeP$_2$	&	&	9.33	&	8.81	&	&	0.85	&	&	0.87	&	1.61	& {\bf{1.54}} &	1.58$^c$	&	0.90&-0.01	\\
BeSiAs$_2$	&	&	9.76	&	9.32	&	&	0.82	&	&	0.97	&	1.82	& 1.79 &	{\bf{1.33}}$^c$	&	1.11&0.09	\\
BeSiP$_2$	&	&	8.63	&	8.37	&	&	0.86	&	&	1.18	&	1.95	& 1.90 &	{\bf{1.75}}$^c$	&	1.30&0.03	\\
BeSnAs$_2$	&	&	11.52	&	10.79	&	&	0.79	&	&	0.56	&	1.32	& 1.34 &	{\bf{1.25}}$^c$	&	1.15&0.11	\\
BeSnP$_2$	&	&	9.57	&	9.23	&	&	0.82	&	&	0.88	&	1.60	& 1.57 &	{\bf{1.78}}$^c$	&	1.98&0.03	\\
CdGeAs$_2$	&	&	25.26	&	12.92	&	&	0.74	&	&	0.11	&	{\bf{0.43}}	& 0.17&	0.26$^c$	&	0.57&0.04	\\
CdGeP$_2$	&	&	11.45	&	9.55	&	&	0.78	&	&	0.63	&	1.39	& 1.38 &	{\bf{1.61}}$^c$	&	1.72&0.00	\\
CdSiAs$_2$	&	&	15.30	&	10.64	&	&	0.76	&	&	0.33	&	1.12	& 1.19 &	{\bf{1.29}}$^c$	&	1.55&0.04	\\
CdSiP$_2$	&	&	9.55	&	8.94	&	&	0.81	&	&	1.42	&	{\bf{2.09}}	& 2.06 &	1.91$^c$	&	2.20&0.00	\\
CdSnAs$_2$	&	&	13.24	&	4.03	&	&	0.71	&	&	0.07	&	{\bf{0.18}}	& 0.03 &	0.16$^c$	&	0.26&0.03	\\
CdSnP$_2$	&	&	18.53	&	9.34	&	&	0.74	&	&	0.24	&	0.97	& 0.98 &	{\bf{1.10}}$^c$	&	1.17&0.02	\\
MgGeAs$_2$	&	&	11.54	&	9.31	&	&	0.72	&	&	0.49	&	{\bf{1.34}}	& 1.26 &	1.27$^c$	&	1.60&0.05	\\
MgGeP$_2$	&	&	8.48	&	7.75	&	&	0.75	&	&	1.52	&	2.26	& {\bf{2.15}} &	2.28$^c$	&	2.17&-0.03	\\
MgSiAs$_2$	&	&	9.14	&	8.47	&	&	0.74	&	&	1.21	&	{\bf{2.02}}	& 1.91 &	1.40$^c$	&	2.00&0.04	\\
MgSiP$_2$	&	&	7.79	&	7.42	&	&	0.77	&	&	1.37	&	{\bf{2.18}}	& 2.03 &	1.83$^c$	&	2.26&0.00	\\
MgSnAs$_2$	&	&	12.13	&	9.02	&	&	0.69	&	&	0.31	&	{\bf{1.17}}	& 1.10 &	1.07$^c$	&	1.20&0.07	\\
MgSnP$_2$	&	&	8.21	&	7.52	&	&	0.72	&	&	1.18	&	2.02	& 1.93 &	{\bf{2.05}}$^c$	&	2.48&0.00	\\
ZnGeAs$_2$	&	&	$-$	&	11.59	&	&	0.77	&	&	0.06	&	{\bf{0.72}}	& 0.65 &	0.67$^c$	&	1.15&0.01	\\
ZnGeP$_2$	&	&	10.47	&	9.24	&	&	0.81	&	&	1.14	&	1.97	& {\bf{1.91}} &	2.25$^c$	&	1.80&0.01	\\
ZnSiAs$_2$	&	&	11.59	&	10.16	&	&	0.79	&	&	0.81	&	{\bf{1.76}}	& 1.79 &	1.94$^c$	&	1.60&0.05	\\
ZnSiP$_2$	&	&	9.34	&	8.68	&	&	0.83	&	&	1.35	&	{\bf{2.09}}	& 2.04 &	1.92$^c$	&	2.30&0.01	\\
ZnSnAs$_2$	&	&	$-$	&	11.69	&	&	0.74	&	&	0.00	&	{\bf{0.46}}	& 0.44 &	0.44$^c$	&	0.75&0.00	\\
ZnSiP$_2$	&	&	10.45	&	8.68	&	&	0.83	&	&	1.35	&	{\bf{2.09}}	& 1.48 &	1.92$^c$	&	2.30&0.01	\\
\hline\hline
TME&&&&&&&0.95&{\textbf{0.02}}&0.15&0.17\\
TMAE&&&&&&&0.95&{\textbf{0.24}}&0.27&0.28\\
TMARE&&&&&&&0.57&{\textbf{0.17}}&0.20&0.19\\
\hline\hline
\end{tabular}
\begin{flushleft}
a)  Spin-orbit coupling is defined as $\Delta_{SOC}$=E$_g^{PBE}-$E$_g^{PBE-SOC}$ which has been added with all SOC un-corrected bandgaps obtained from DFT calculations.\\
b) Present work calculated with 
$GW_0$ from VASP. See main text for details.\\
c) $GW_0$ values from ref.~\cite{shaposhnikov2012abinitio}. These calculations are performed using VASP code.\\
d) $G_0W_0$@PBE from ref.~\cite{Zhang2013near}.\\
e) $G_0W_0$@PBE+U from ref.~\cite{Zhang2013near}.
\end{flushleft}
\end{table*}
\begin{figure*}
\begin{center}
\includegraphics[width = 20 cm]{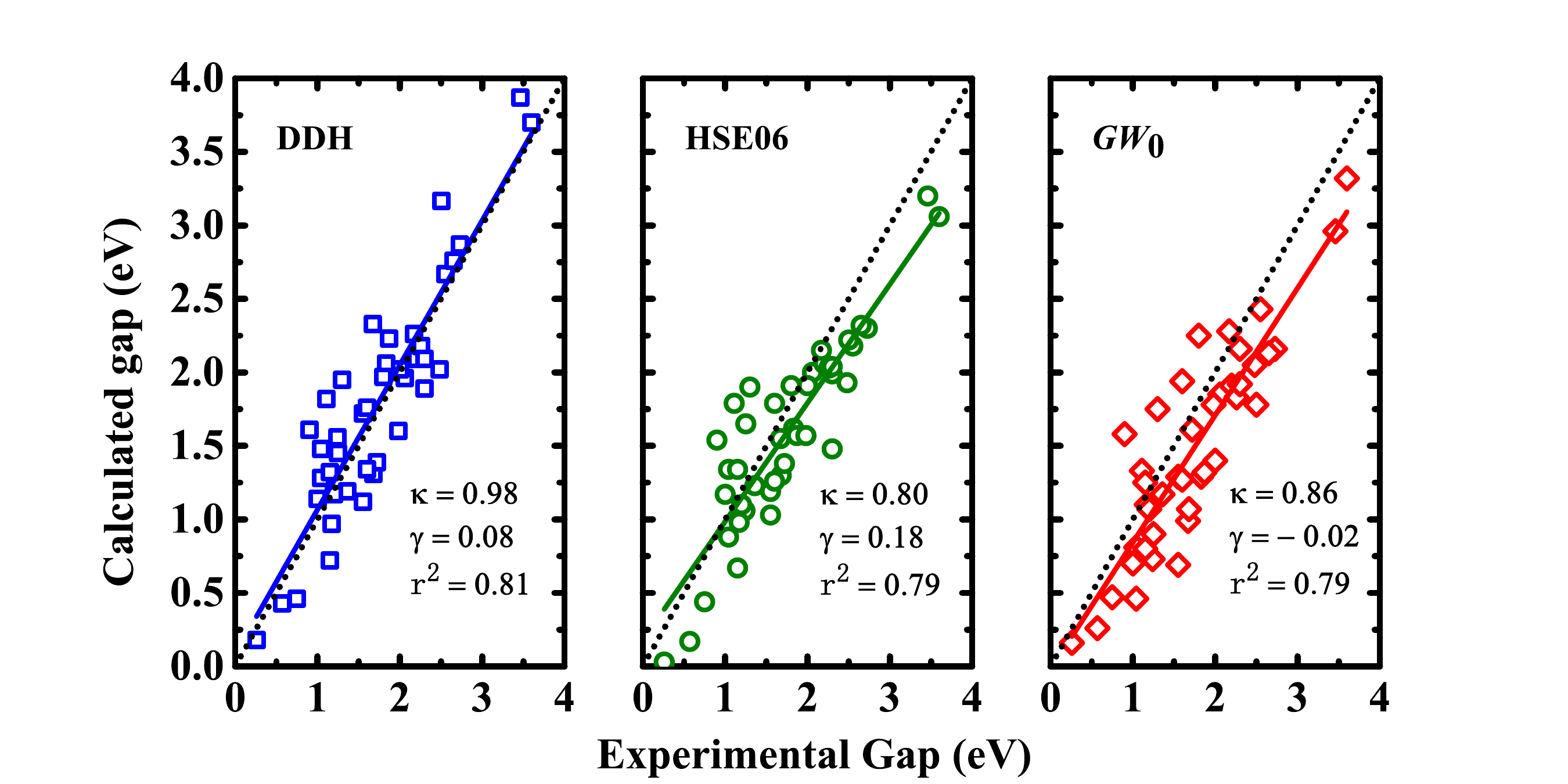}
\end{center}
\caption{\label{fig4} The calculated versus experimental band gaps of $42$ chalcopyrite semiconductor using DDH, HSE06, and $GW_0$. The linear regressions formula is defined as, E$_g^{calc}$ = $\kappa$E$_g^{expt}$ + $\gamma$, where $\kappa$ is the slope, $\gamma$ is the intercept (in
eV), and $r^2$ is the correlation coefficient. The dashed lines in every panel give E$_g^{calc}$ = E$_g^{expt}$. For $GW_0$, use the best value from the Table~\ref{tab-bg}.}
\end{figure*}
\begin{figure}
\begin{center}
\includegraphics[width=\columnwidth]{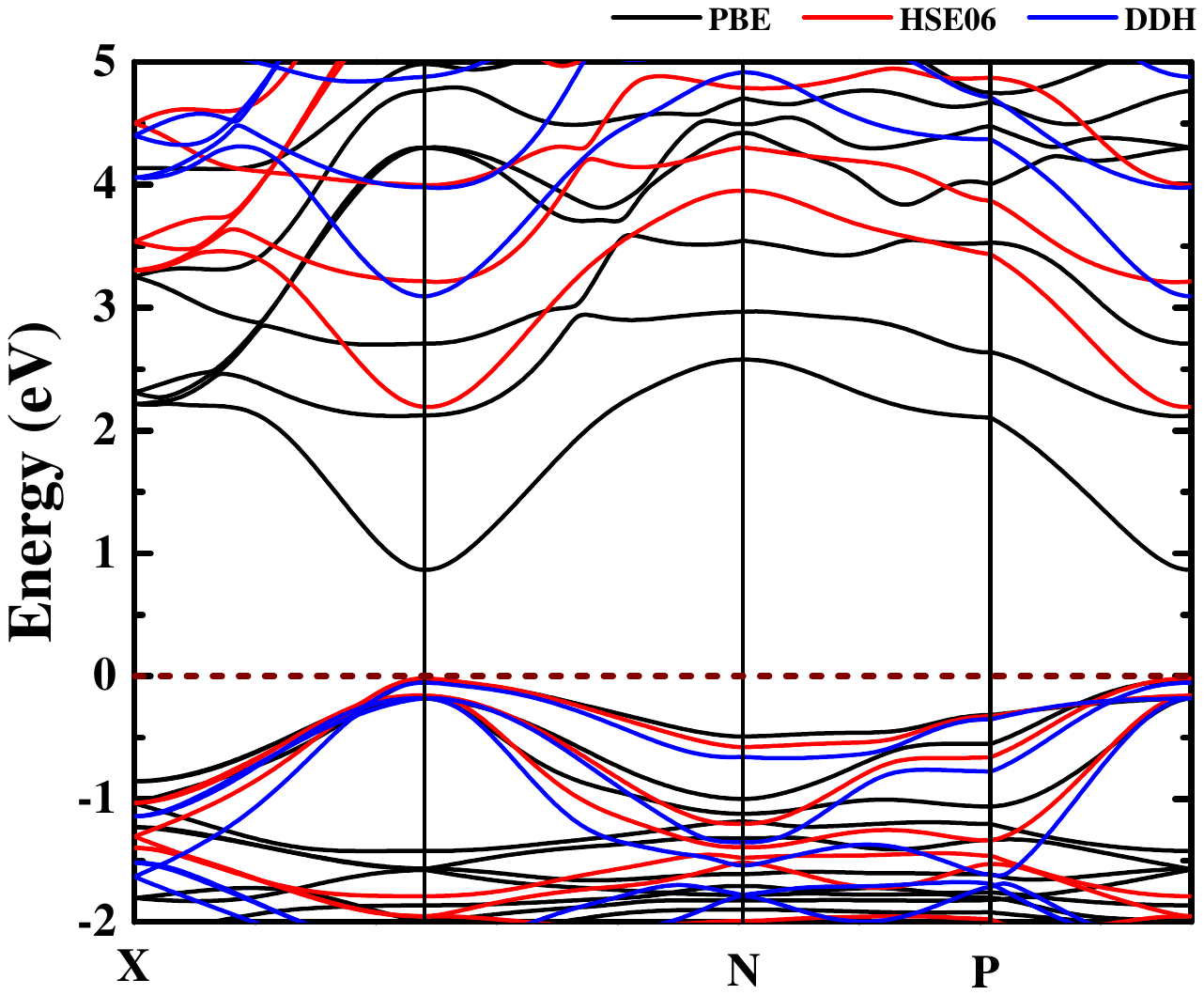}
\end{center}
\caption{\label{band-structure} Bandstructure of CuGaS$_2$ calculated using PBE, HSE06 and DDH. The dashed line indicates the Fermi energy.}
\end{figure}
\begin{figure}
\begin{center}
\includegraphics[width=\columnwidth]{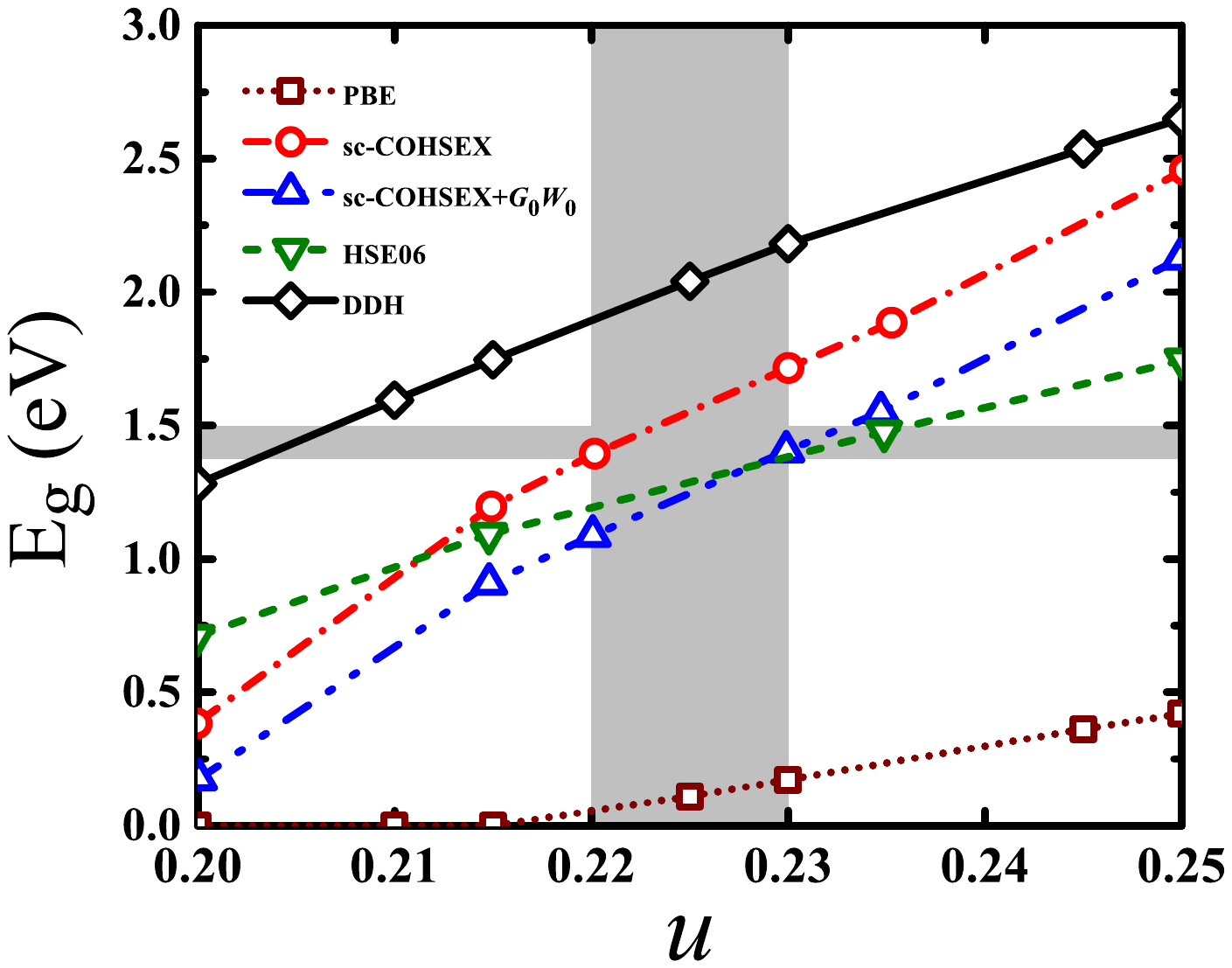}
\end{center}
\caption{\label{u-vs-eg} Band gap vs the anion
displacement $u$ for CuInS$_2$ as obtained using different methods. The  PBE, HSE06, sc-COHSEX,
and sc-COHSEX+$G_0W_0$ values are taken from ref.~\cite{vidal2010strong}.}
\end{figure}
\begin{figure}
\begin{center}
\includegraphics[width=\columnwidth]{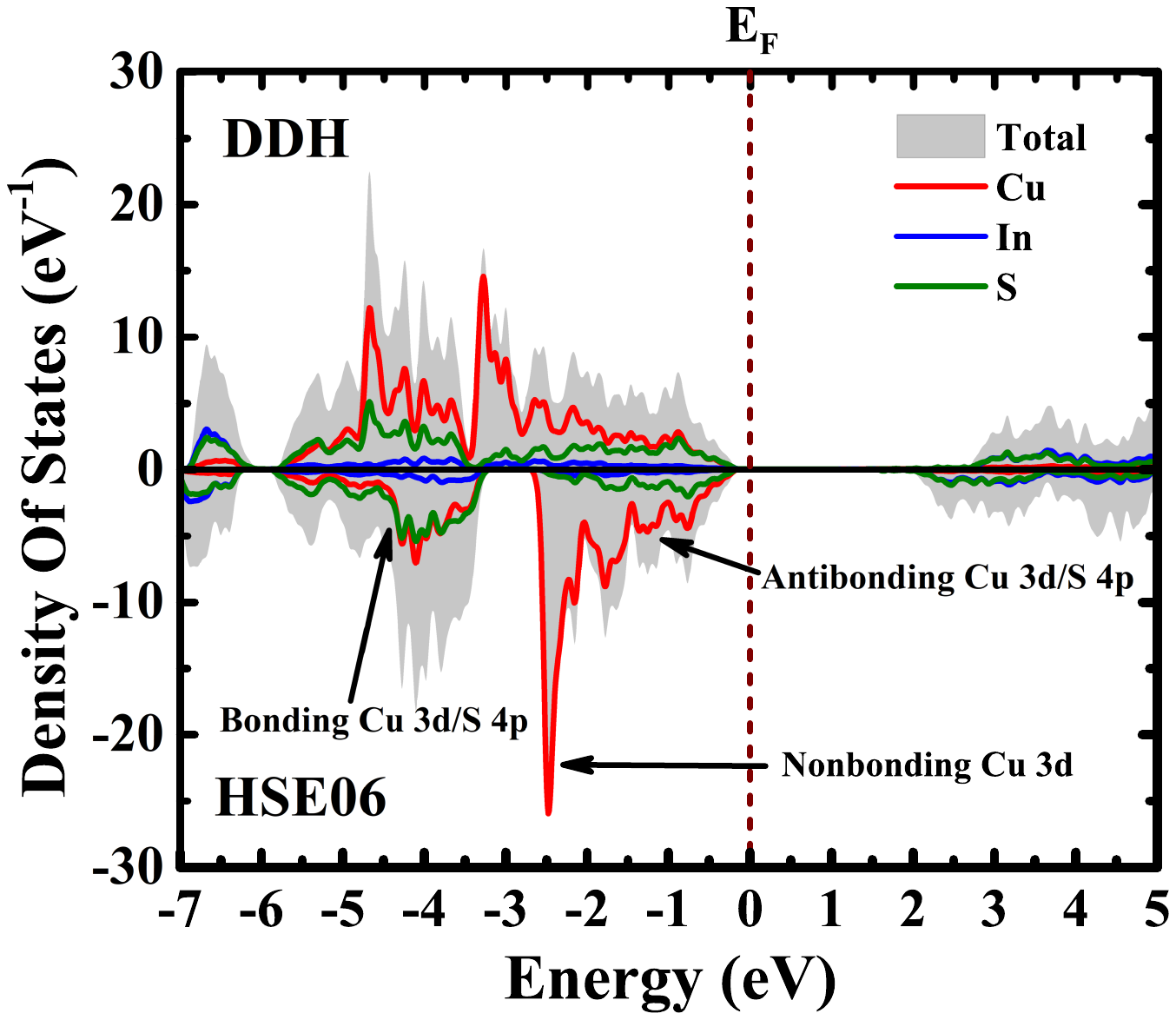}
\end{center}
\caption{\label{dos} Density of states (DOS) for CuInS$_2$ calculated using DDH and HSE06.}
\end{figure}
\begin{center}
\begin{figure}
\includegraphics[width=10 cm]{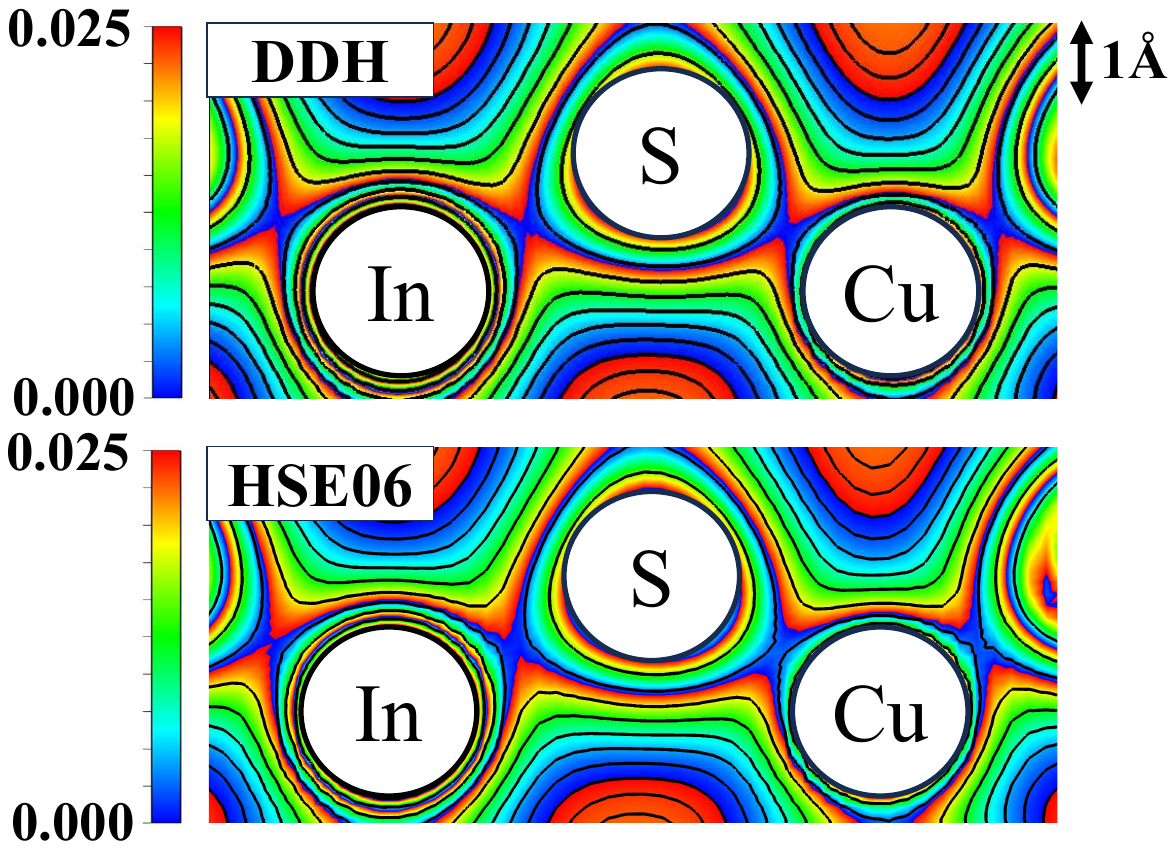}
\caption{\label{chg-density} Electronic charge density distribution contours of CuInS$_2$ as obtained from DDH and HSE06. The positions of Cu, In, and S atoms are shown.  The logarithmic scale is used for the better visualization of
the isosurfaces.}
\end{figure}
\end{center}

\begingroup
\begin{table}
\caption{\label{d-band} Mean positions of the occupied $d$ band (in eV) relative to the VBM for selective semiconductors. The theoretical values are calculated by averaging the $d$ state
energies at the $\Gamma$ point.}
\begin{tabular}{cccccccccccccccccccccccccccccccccc}
\hline\hline
	Solids&&& PBE&&&DDH&&&HSE06&&&$GW_0$ \\
	\hline
AgGaS$_2$&&&-14.8&&&-19.8&&&-17.1&&&-17.3\\
AgInS$_2$&&&-14.4&&&-17.3&&&-15.9&&&-16.6\\
AgInTe$_2$&&&-14.8&&&-17.8&&&-16.4&&&-16.9\\
AgGaTe$_2$&&&-15.3&&&-20.4&&&-17.9&&&-16.7\\
AgGaSe$_2$&&&-14.9&&&-20.0&&&-17.3&&&-16.5\\
CuGaS$_2$&&&-15.4&&&-20.8&&&-17.6&&&-18.1\\
CuInS$_2$&&&-15.0&&&-17.7&&&-16.5&&&-17.0\\
CuGaTe$_2$&&&-15.4&&&-21.1&&&-17.9&&&-18.5\\
\hline\hline
\end{tabular}
\begin{flushleft}
\end{flushleft}
\end{table}

\subsection{Analysis of band structures}

\subsubsection{Bandgaps and band structures}
Now, let's delve into the performance of DDH in estimating the band gaps. Table~\ref{tab-bg} provides a comprehensive comparison by presenting (g)KS band gaps obtained from PBE, HSE06, and $GW_0$. Band gap is defined as $E_g=\epsilon_{LUMO}-\epsilon_{HOMO}$, where $\epsilon_{LUMO}$ and $\epsilon_{HOMO}$ represent the corresponding lowest unoccupied molecular orbital (LUMO) and highest occupied molecular orbital (HOMO) eigenvalues.

{\textcolor{black}{Inspecting the band gaps of I-II-VI$_2$, one can readily observe that, as usual, PBE underestimates the band gaps for all systems and becomes zero for AgGaTe$_2$, AgInTe$_2$, CuGaSe$_2$, CuGaTe$_2$, and CuInS$_2$. Although HSE06 offers an improvement over PBE, yet important improvement is observed in DDH calculations.
Using DDH, the band gap increases to $\sim~0.8$ eV for Ag-based chalcopyrites compared to HSE06, making DDH band gaps close to the experimental values.
Similarly, for Cu-based chalcopyrites, the gaps obtained using DDH are also in good agreement. We observe good performance for CuAlSe$_2$, CuGaTe$_2$, and CuInTe$_2$ using DDH, which brings it closer to experimental values. One may note that for Cu-based chalcopyrites, the interplay between Cu $d$ and anions $p$ is identified as a crucial factor~\cite{Zhang2013near,vidal2010strong,Irene2011first}.}}
Interestingly, for CuGaS$_2$, CuGaSe$_2$, CuInS$_2$, and CuInSe$_2$, the underestimation in the band gap from $G_0W_0$ and $GW_0$ is noticeable. As mentioned in ref.~\cite{shaposhnikov2012abinitio}, the screening is not treated correctly by $G_0W_0$ and the PBE+U method may be the better
starting point for quasiparticle $GW$ calculations~\cite{Zhang2013near,vidal2010strong,Irene2011first}.
For example, using the PBE+U as a starting point for $G_0W_0$, the band gaps of CuInS$_2$ improve to 1.41 eV, where $G_0W_0$@PBE value was $0.77$ eV (in our present calculation it is $0.69$ for $GW_0$ which is close to $G_0W_0$@PBE value of ref.~\cite{Zhang2013near}).
A similar improvement in results is also
noticed when employing
PBE+U as a starting point for Cu-based chalcopyrite semiconductors.
In accordance with the insights from ref.~\cite{vidal2010strong}, these systems call for careful treatment of $GW$, requiring fully self-consistent (sc)COHSEX or $GW$ for precise band gap predictions. Nevertheless, the remarkable performance of DDH in these semiconductor materials hints at its potential to be the preferred method, offering a commendable balance between accuracy and relatively lower computational cost compared to the self-consistent $GW$. {\textcolor{black}{Similarly, DDH also delivers accurate results for II-IV-V$_2$ semiconductors and is closely aligned with the experimental values.}}

{\textcolor{black}{Fig.~\ref{fig4} shows the experimental versus theoretical band gaps obtained using DDH, HSE06, and $GW_0$. We use linear regression analysis to understand better the errors coming from different methods. We calculate slope ($\kappa$), interception ($\gamma$), and correlation coefficient ($r^2$) for comparison. For DDH, the $\kappa$, $\gamma$, and correlation coefficient $r^2$ are found to be about $0.98$, $0.08$, and $0.81$ which are slightly better than  HSE06. Considering the $GW_0$, its error statistics are similar to HSE06. The most relevant parameter here is the $r^2$, which is marginally better for DDH
functional than HSE06 and GW0. Also, as shown in Table~\ref{tab-bg}, in terms of total mean absolute error (MAE) and mean absolute relative error (MARE), DDH performs marginally better than HSE06 as well as $GW_0$.}}

In Fig.~\ref{band-structure}, we plot the band structures from PBE, HSE06, and DDH for CuGaS$_2$, a direct band gap (located at $\Gamma-$ point) semiconductor. We observe an identical band structure and curvature apart from the shift in the conduction band for different methods.
Specifically, DDH shows a noticeable $\sim 0.97$ eV shift in the conduction band at the $\Gamma$-point compared to HSE06.
\subsubsection{Variation of bandgaps with $u$}
To understand the variation of the band gaps with the distortion parameter `$u$' (defined in Eq.~\ref{eqq1}),
, we show the band gap variation of CuInS$_2$, one of the prototype chalcopyrite semiconductor, using various methods in Fig~\ref{u-vs-eg}. This is important because, experimentally, one can observe the stability of chalcopyrite with the variation of $u$. 
Typically, PBE predicts the system to be metallic for $0.20\leq u \leq 0.215$. HSE06 improved over PBE but underestimated bandgap by $\sim 0.2$ eV. A very similar performance is also observed from various sc-COHSEX+$G_0W_0$, while fully self-consistent sc-COHSEX improves over single shot sc-COHSEX, i.e., sc-COHSEX+$G_0W_0$~\cite{vidal2010strong}. In contrast, the improvements in the bandgaps for DDH are quite noticeable over the ranges of $u$. Although DDH overestimates the bandgaps only by $\sim 0.2$ eV compared to the experimental values, it is well within the shaded part of Fig.~\ref{u-vs-eg}. For each $u$, the DDH uses different Fock mixing by $\epsilon_\infty^{-1}$. The seemingly different band gap values from different methods occur because of the different screening of these systems with various $u$, which is expected due to
interplay between the $u$ and the hybridization of the $p-d$ orbitals~\cite{vidal2010strong}.

\subsubsection{Analysis of DOS and charge density}

More analysis can also be drawn from the density of states (DOS) for CuInS$_2$ in Fig~\ref{dos}. Typically, the main contribution comes from metal Cu$-3d$, anion S$-3p$, and their (non-bonding/anti-bonding) hybridization. For DDH, there is a downward shift of non-bonding $3d$ states because of the stronger hybridization, which reduces the $p-d$ repulsion, hence the enlargement of bandwidth.

The differences in the
performances of DDH and HSE06 can also be drawn from the charge density contour plot of CuInS$_2$ as shown in Fig~\ref{chg-density}. As known, the In$-$S bond is ionic, whereas Cu$-$S
bond is covalent. The nature of covalency changes due to the repulsive $p-d$ non-bonding nature, which is depicted through the change density iso-surface plot. The reduction in the isosurface value between Cu$-$S indicates a decrease of $p-d$ repulsion in DDH; hence, the atomic distance becomes slightly lower than HSE06.

\subsubsection{Positions of valance $d-$ bands}

Finally, in Table~\ref{d-band}, we show the mean positions of the occupied $d$ band (in eV) (relative
to the VBM) for several chalcopyrite semiconductors. As PBE functional suffers from the known de-localization, its occupied $d$ band is quite underestimated. Although HSE06 improves over PBE, DDH generally recovers
the mean positions of the occupied $d$ band quite remarkably compared to both HSE06~\cite{jana2023simple}. However, our results suggest non-empirical DDH obtains a bit of deep occupied $d-$ band positions compared with $GW_0$. This is because, for these systems, the band gaps are overestimated from DDH. 
Unfortunately, no experimental results are available to compare with.

The analysis of band gaps, distortion parameter versus band gap variations, the density of states, and charge density indicate the screening effect as determined using DDH, which is important for determining accurate properties from hybrid functionals. Importantly, a judicious choice of the percentage of the Fock screening is required for a more accurate prediction of the properties. In that case, the functional becomes empirical or tuned DDH~\cite{OhadWingGant2022}. However, looking at the nonempirical settings and superiority of the obtained properties, the present DDH can be considered as one of the useful methods, especially when higher-level accurate methods like $GW$ are necessary but computationally unaffordable.

\begin{figure}
    \centering
    \includegraphics[width=\columnwidth]{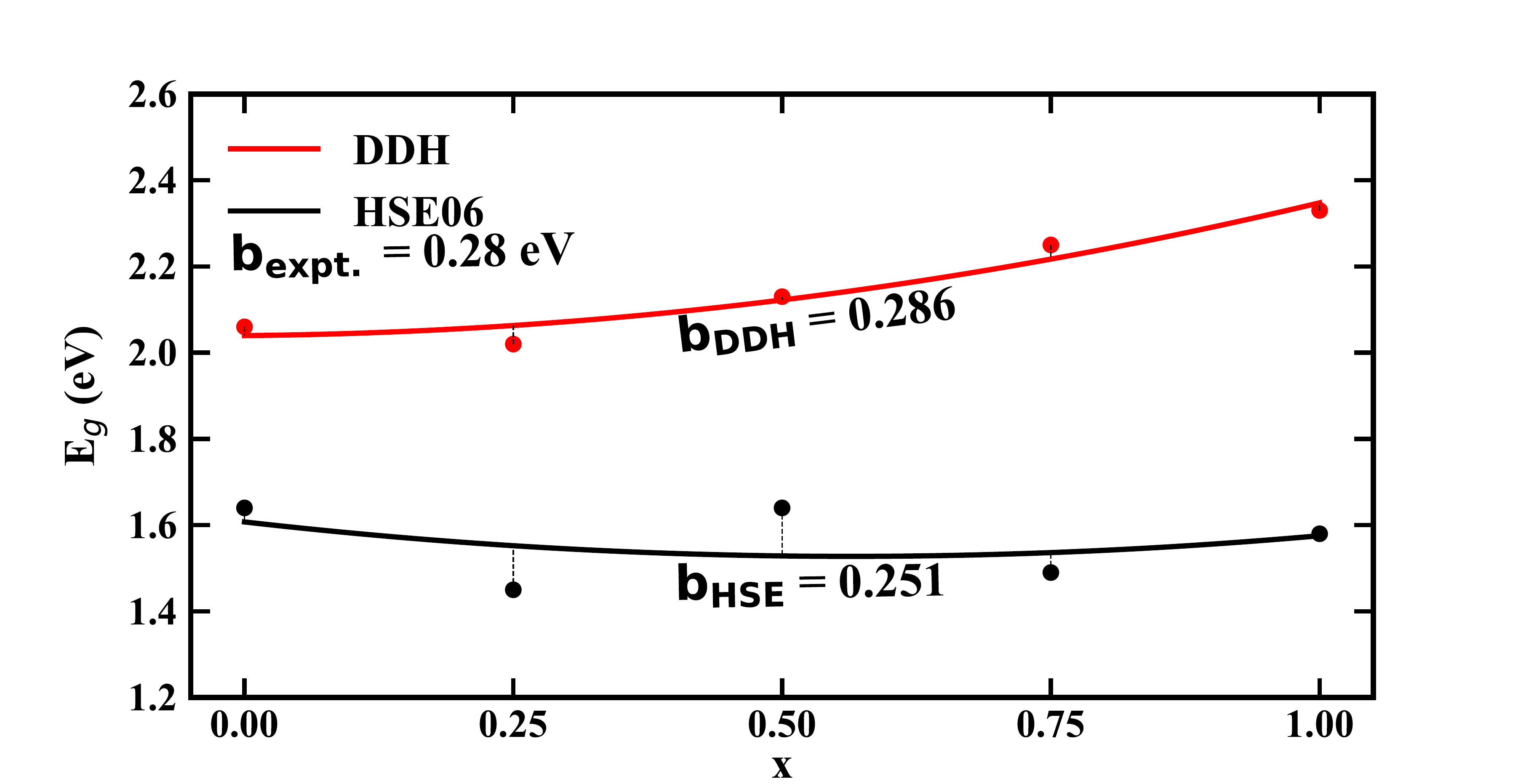}
    \caption{The band gap values for Cu$_{x}$Ag$_{1-x}$GaSe$_{2}$ alloy are calculated at compositions $x=0.0, 0.25, 0.50, 0.75, \text{ and } 1.0$, with DDH and HSE06. The curves are fitted using the Eq.~\ref{opt.eq}.}
  \label{bowing}
\end{figure}

\subsection{Optical bowing parameters}

The band gap of ternary chalcopyrites varies from CuBX$_{2}$ to AgBX$_{2}$ due to the different sizes of Cu and Ag atoms. This variation is evident in Table~\ref{tab-bg} for all XC approximations. The band gap varies  as the Cu/Ag ratio changes~\cite{shaukat1990composition}. These size-dependent variations are further intensified by band bowing, known as optical bowing~\cite{mourad2012theory}. This effect arises from atomic-level fluctuations in the lattice structure of (Cu, Ag)BX$_2$ compounds~\cite{mudryi2003optical}. Due to level repulsion between chalcopyrite energy levels in the alloy, the band gap of the alloy experiences a downward shift from the linear average, as described by the following equation:
\begin{equation}
    E_{g}(A_{x}B_{1-x}) = xE_{g}(A) + (1-x)E_{g}(B) - bx(1-x)~,
    \label{opt.eq}
\end{equation}
where $E_{g}(A)$ and $E_{g}(B)$ are the band gaps of A and B for the compound $A_xB_{1-x}$.
Considering the quaternary chalcopyrite semiconductor alloy Cu$_{x}$Ag$_{1-x}$GaSe$_{2}$, the common Ga$-$X bond length remains almost unchanged with concentration $x$, while the Ag$-$Se and Cu$-$Se bond lengths associated with Ag and Cu increase with $x$~\cite{wei1995band}. This is consistent with the nearly identical Ga-Se bond lengths in CuGaSe$_2$ and AgGaSe$_2$ because the local environment of Ga does not change with the alloy concentration $x$. In the case of DDH, the band gap varies from 2.33 to 2.06 eV as we go from CuGaSe$_2$ to AgGaSe$_2$, and in the case of HSE06, it varies from 2.55 to 2.62, as shown in Table~\ref{tab-bg}. The optical bowing parameter '$b$' can be obtained using the band gap difference. The band gaps for Cu$_{x}$Ag$_{1-x}$GaSe$_{2}$ with concentrations of ($x=$0.0, 0.25, 0.50, 0.75, 1.0) are calculated using PBE, DDH and HSE06 XC approximations. The band gap value using PBE for $x=$0.0 and 0.25 is 0 eV, so PBE band gap errors are largely canceled in the calculation. The optical bowing parameter '$b$' can be calculated at different $x$ along with pure AgGaSe$_2$ and CuGaSe$_2$ according to Equation~\ref{opt.eq}.

Figure~\ref{bowing} is plotted by curve-fitting the calculated band gap using HSE06 and DDH using Eq.~\ref{opt.eq}. The experimental value of '$b$' for Cu$_{x}$Ag$_{1-x}$GaSe$_{2}$ is 0.280 eV ~\cite{choi2000optical}. The value of '$b$' using HSE06 is 0.251 eV, whereas for DDH, it is 0.286 eV. Our study shows that the value of '$b$' using DDH is in very good agreement
with the experimental optical bowing parameter compared to HSE06.

%
\begin{figure}
\begin{center}
\includegraphics[width=\columnwidth]{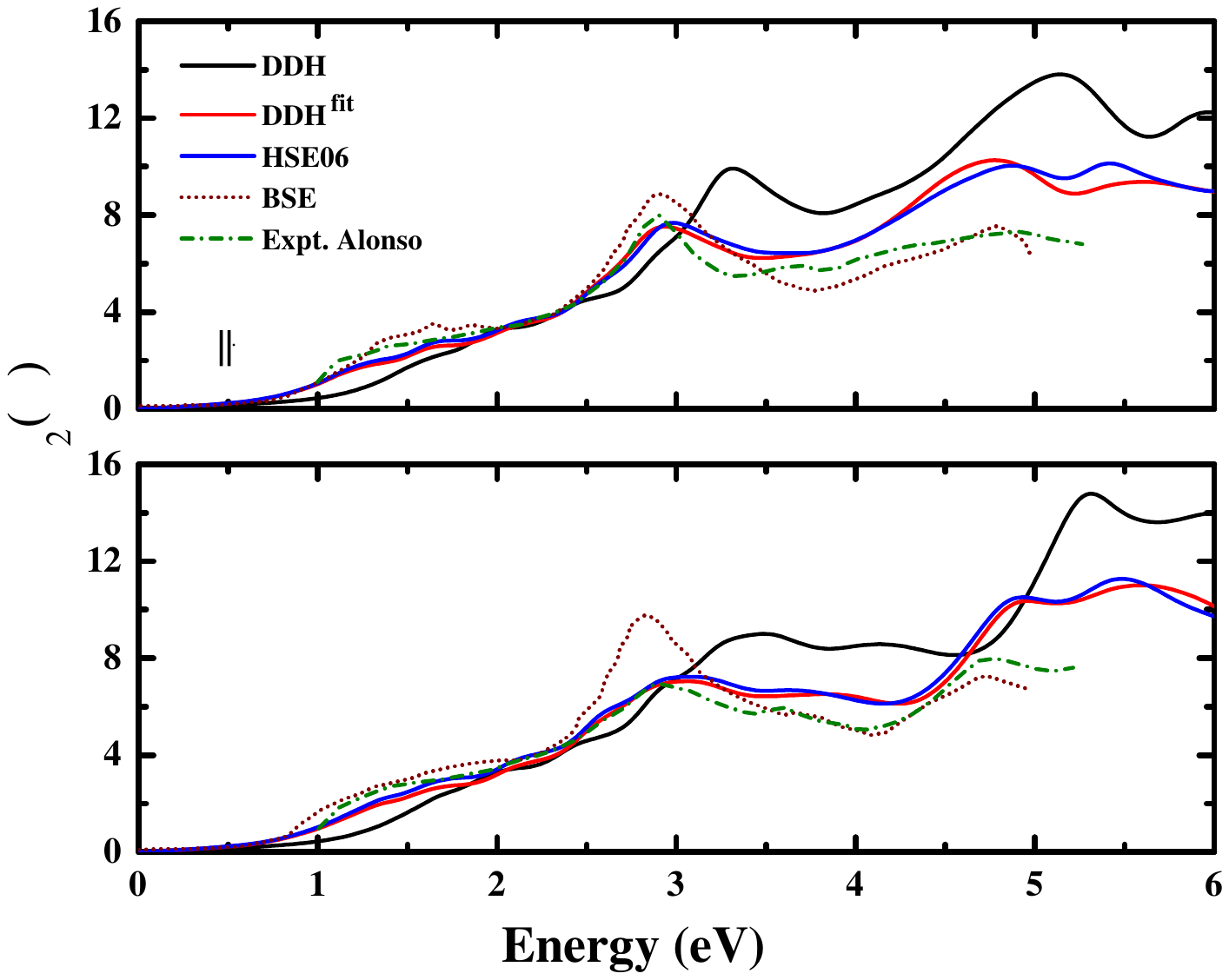}
\end{center}
\caption{\label{fig-cuinse2-opt} Imaginary part of absorption spectra ($\epsilon_2(\omega)$) of CuInSe$_2$ (upper panel) for light polarized along the $c$ axis (upper panel) and for light polarized perpendicular to the $c$ axis (lower panel) calculated with DDH (non-empirical) and DDH$^{fit}$ (with empirical tuning of parameter $\gamma$) and HSE06. The experimental and BSE@$GW$ results are taken from Alonso et al.~\cite{alonso2001optical} and Korbel et al. ref.~\cite{Sabine2015optical}, respectively.}
\end{figure}
\begin{figure}
\begin{center}
\includegraphics[width=\columnwidth]{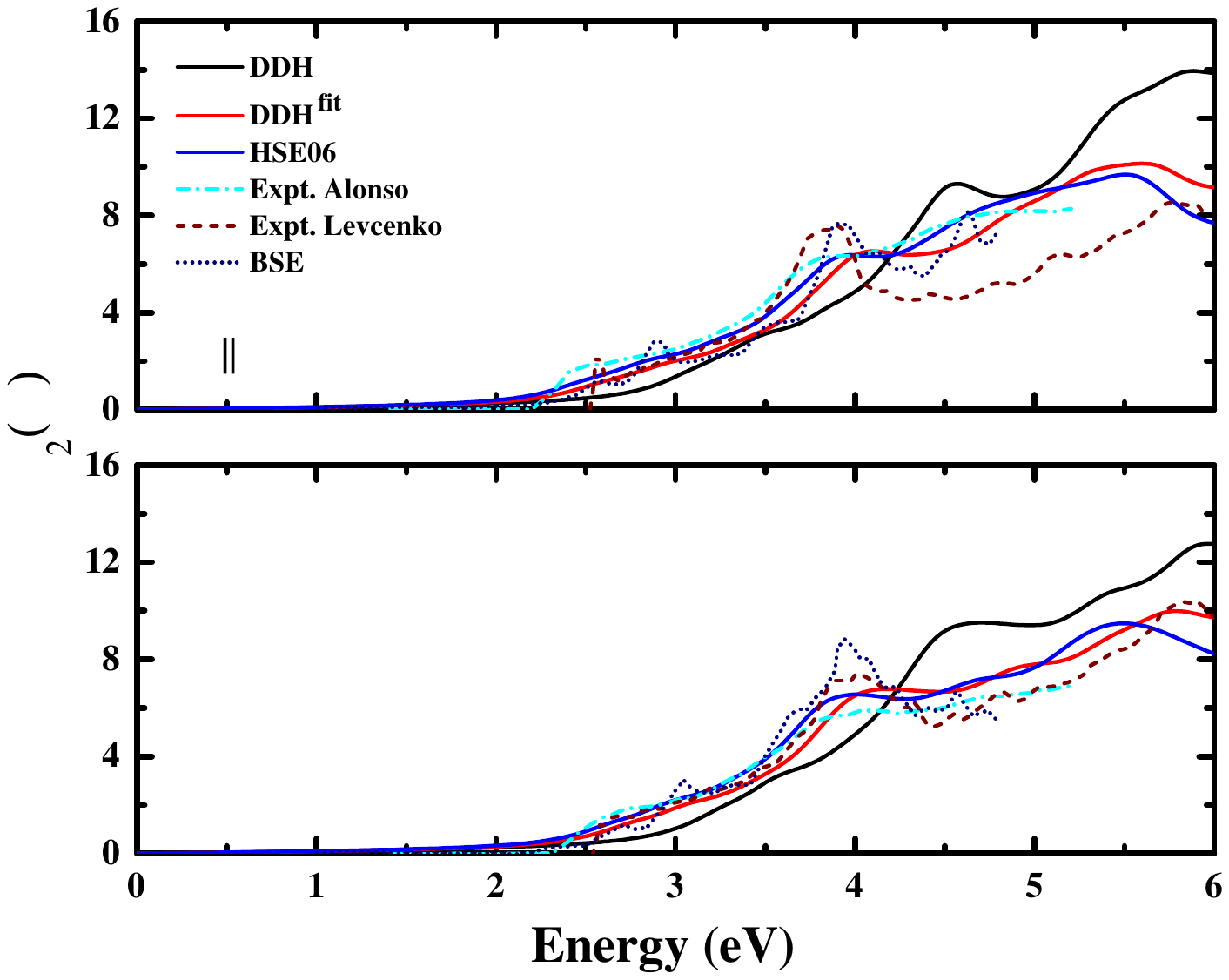}
\end{center}
\caption{\label{fig-cuigas2-opt} Same as Fig.~\ref{fig-cuinse2-opt} but calculated for CuGaS$_2$. 
The experimental values are taken from Alonso et al.~\cite{alonso2001optical} and  Levcenko et al.~\cite{Levcenko2007optical}. The BSE@$GW$ results are taken from  Aguilera et. al.~\cite{Irene2011first}.}
\end{figure}

\subsection{Description of absorption
spectra}
Optical absorption properties of solids in DFT are mostly calculated using the linear-response TDDFT (LR-TDDFT) by solving the Casida
equation using KS or gKS orbitals~\cite{cassida-equation}. However, to get an accurate optical spectrum for a solid, it is necessary to get the correct band gap and accurate treatment of the electrons and holes. {\textcolor{black}{State-of-the art BSE@$GW$ includes the physics correctly but comes with
huge computational expenses.}} {\textcolor{black}{Typically, in VASP, the excitation spectra for both the TDDFT and BSE@$GW$ methods are calculated by solving the Casida/Bethe-Salpeter equation~\cite{cassida-equation,Onida2002electronic,Sander2015beyond,Sander2017macroscopic,tal2020accurate}.}}

Within the Tamm-Dancoff approximation, the TDDFT spectrum is calculated by solving the matrix element of direct transition from occupied to unoccupied states and the electron-hole ($e-h$) interaction as~\cite{cassida-equation,Sander2015beyond,Sander2017macroscopic,tal2020accurate},
\begin{eqnarray}
 A_{ij;ab}=\omega_{ia}\delta_{ij;ab}+\langle ib|K_{Hxc}|aj\rangle~,  \label{casida}
\end{eqnarray}
where $\omega_{ia}=\varepsilon_i-\varepsilon_a$. The occupied states are noted as $i$, $j$, unoccupied states as $a$, $b$, and corresponding KS or gKS eigenvalues are denoted by $\varepsilon$. The matrix element $\langle ib|K_{Hxc}|aj\rangle$ is evaluated using the Hartree plus XC kernel, $K_{Hxc}$, which includes the $e-h$ interaction. Because the semilocal functionals underestimate the band gaps, typically, the first term of the right-hand side of Eq.~\ref{casida}, depending on the KS eigenvalues, is not well treated. The $e-h$ interaction term is further simplified using the framework of the gKS and screened-DDH as Hartree, screened exchange, and
the term involves the XC kernel as~\cite{tal2020accurate},
\begin{eqnarray}
\langle ib|K_{Hxc}|aj\rangle &=& 2\langle ib|V(|{\bf{q+G}}|)|aj\rangle\nonumber\\
&-& \langle ib|f_{xc}^{non-local}|ja\rangle+\langle ib|f_{xc}^{local}|aj\rangle\nonumber\\
\end{eqnarray}
The first term on the right side of the above matrix element involves the Hartree term common for both TDDFT and BSE@$GW$. The term involving $\langle ib|f_{xc}^{non-local}|ja\rangle = \langle ib|\epsilon_\infty^{-1}(|{\bf{q+G}}|)V(|{\bf{q+G}}|)|ja\rangle$ does not appear when considering only the semilocal XC functional. The local XC kernel, $f_{xc}^{local}$, is the functional derivative of semilocal XC approximations with respect to the density and does not include any excitonic effect. However, some specially designed low-cost XC kernels are also developed and include necessary features~\cite{Trevisanutto2013optical,terentjev2018gradient,sharma2011bootstrap,rigamonti2015estimating,van2002ultranonlocality,
cavo2020accurate,Byun2017assessment,Byun2020time}. Importantly, for a short-range screened hybrid like HSE06, $\epsilon_\infty^{-1}(|{\bf{q+G}}|)=\beta(1-e^{-|{\bf{G}}|{^2}/(4\mu)})$ and the resultant screened exchange varies as $\sim\beta$, a constant with $\beta=0.25$ at $q\to 0$. However, for bulk systems, the correct behavior of screened exchange must be $~\sim\frac{\epsilon_\infty^{-1}(|{\bf{G}}|)}{ q^2}$ at $q\to 0$, which is the key to improving the optical absorption spectra from DDH~\cite{Ullrichtddft,paier2008dielectric,Yang2015simple,Wing2019comparing,stadele1999exact,petersilka1996excitation,kim2002excitonic,Sun2020optical,sun2020lowcost,Kootstra2000application},
where the high-frequency dielectric constant of the material replaces dielectric function~\cite{tal2020accurate}.

{\textcolor{black}{On the other hand, in BSE, the same set of equations are solved~\cite{Onida2002electronic,tal2020accurate}. However, the orbitals are related to the previous $GW$ calculations~\cite{Onida2002electronic,tal2020accurate}, typically obtained from different levels of approximations, where the screened exchange is frequency-dependent (through the dielectric function)~\cite{tal2020accurate} and the inclusion of the ``nanoquanta'' vertex correction may also required for accurate calculations~\cite{Shishkin2007accurate}. For the corresponding VASP implementation of the BSE equation and details differences of BSE and TDDFT using DDH, the readers are referred to ref.~\cite{tal2020accurate}. 
While BSE calculations neglect dynamical effects, it's worth noting that determining the screened exchange in prior $GW$ steps adds complexity to the overall problem~\cite{tal2020accurate}.
}}

Finally, in TDDFT, the frequency-dependent and small wave vectors limit of the imaginary ($\epsilon_2(\omega)$) part of the macroscopic dielectric function (optical), $\epsilon^M$ is calculated via
\begin{eqnarray}
 \epsilon_2(\omega)&=& \Im \{\lim_{q\rightarrow 0}\epsilon^M(q,\omega)\}~,
\label{eq8bb}
\end{eqnarray}
where $\lim_{q\rightarrow 0}\epsilon^M(q,\omega)$ is given by the Eq. (48) of ref.~\cite{Sander2017macroscopic}, which is also related to the calculated matrix elements of Eq.~(\ref{casida}) and how the kernels are evaluated. It is quite apparent that the results obtained from TDDFT using the semilocal only and hybrid DFT functionals give drastically different results.

In the following, we choose  CuInSe$_2$ and CuGaS$_2$ to calculate the TDDFT spectrum for which high-level calculations and experimental values are available. For CuInSe$_2$,  we consider $a=5.780$ \AA, $c=11.618$ \AA, and $u=0.230$ according to the experimental values in Table I of the ref~\cite{kim2016screened}. Similarly, for CuGaS$_2$ we choose $a=5.351$ \AA, $c=10.478$ \AA, and $u=0.259$ according to the experimental values as supplied in Table II of the ref~\cite{Han2017defect}. To make a meaningful comparison, we compare the results from the KS (as obtained using DDH and HSE06) applying TDDFT, BSE solving the $GW$ (BSE@$GW$), and the experimental dielectric functions. Fig.~\ref{fig-cuinse2-opt} presents the absorption spectra for CuInSe$_2$ and Fig.~\ref{fig-cuigas2-opt} for CuGaS$_2$. In both cases, we show the spectra for the light-polarized
perpendicularly to the $c$ axis (($\epsilon_2^{xx}(\omega)+\epsilon_2^{yy}(\omega)$)/2) or along the $c$ axis ($\epsilon_2^{zz}(\omega)$).

Fig.~\ref{fig-cuinse2-opt} compares the absorption onset of TDDDH, TDHSE06, BSE@$GW$, and experimental in the case of CuInSe$_2$. As evident from the figure, a fairly good agreement of TDDDH with BSE@GW (BSE@$GW$ spectrum is taken from ref.~\cite{Sabine2015optical}) and experimental spectrum from Alonso et al.~\cite{alonso2001optical} is observed. Notably, excitonic peak positions obtained from TDDDH slightly at higher energies or right-shifted when compared with the experimental and BSE@$GW$. This is because of the overestimations in the orbital energies and hence the band gap values from DDH (connected to the Casida Eq.~\ref{casida}). On the other hand, in the case of TDHSE06, we also observe good agreements with TDDDH. Note that the HSE06 absorption spectrum for these semiconductors is also reasonably good compared to the experimental. On the other hand, the advantage of TDDDH is that very accurate spectra can be obtained for both the semiconductor and insulators~\cite{jana2023simple}.

Very similar tendencies are obtained from TDDDH when compared for CuGaS$_2$. As referred to the ref.~\cite{Irene2011first}, for CuGaS$_2$, two absorption spectra are available from Alonso et. al.~\cite{alonso2001optical} and Levcenko et. al.~\cite{Levcenko2007optical}. We consider the experimental spectra of Alonso et al.~\cite{alonso2001optical}, shown in Fig.~\ref{fig-cuigas2-opt}, and those have better agreement with BSE spectra~\cite{Irene2011first}. Considering TDDDH spectra, the first peak is higher energy than the experimental and BSE@$GW$, as shown in Fig.~\ref{fig-cuigas2-opt}. Regarding the TDHSE06, the excitonic peaks also agree with the experimental one, similar to the previous studies.

{\textcolor{black}{One can obtain a good absorption spectrum from TDDDH by empirically tuning the parameter $\gamma$ of Eq.~\ref{hy-eq-gen}. Several other works have adopted this strategy~\cite {Wing2019comparing,OhadWingGant2022,Gant2022optimally,Camarasa2023Transferable}. The following strategies can be adopted to obtain a reasonable spectrum from TDDDH: (i) Tune parameter $\gamma$ of Eq.~\ref{hy-eq-gen} to match with the band gaps of $G_0W_0$ or $GW_0$ when no experimental band gap is available, or (ii) if experimental band gaps are available then the tuning of the $\gamma$ of Eq.~\ref{hy-eq-gen} can be done to match with the experimental band gap of the system, keeping screening parameter $\mu$ fixed. Here, we consider the second strategy. After tuning $\gamma$ from experimental band gaps, we compare the excitation spectrum with the BSE@$GW$ or experimental. We obtain $\gamma=0.08$ for CuGaS$_2$ and $\gamma=0.06$ for CuInSe$_2$ to match the experimental band gaps. 
Consistent agreement is noted in absorption onsets, excitonic peak positions, and the higher energy spectrum when compared to the spectrum for light polarized in both directions. Notably, the tuning procedure is usually not needed for systems where DDH adequately describes band gaps.}}

\section{Conclusions}

{\textcolor{black}{The comparative assessment of the screened-range-separated hybrids (such as HSE06), screened DDH, and methods based on the many-body perturbation theory are assessed for various properties of chalcopyrite semiconductors such as band gaps, optical bowing parameters, and optical absorption spectrum. It is demonstrated that the screened DDH
approach is promising not only as a cheaper alternative to many-body perturbation theory based approaches (such as $G_0W_0$ or $GW$ and BSE@$GW$) but more flexible and physically sound than HSE06. The important fact is that in the screened DDH, the amount of screening correlation is determined from the static dielectric constant instead of a fixed screening used in HSE06. This makes the screened DDH more flexible, especially for the band gaps of chalcopyrites, where the amount of $p-d$ hybridization is determined from screening correlation.
Though the overall mean absolute error suggests that the band gap performance of HSE06, screened DDH, and $GW_0$ (or $G_0W_0$) are quite similar, the screened DDH has a better overall slope, intercept, and correlation coefficient when compared with the experimental band gaps. We also observe that in Cu-based chalcopyrites, the accuracy of screened DDH for band gaps is slightly better than $G_0W_0$@PBE, which strongly depends on the initial starting point.}}

Also, a notable success of the TDDDH in the case of calculating the optical absorption spectrum is demonstrated, which is both cost-efficient and free from empiricism when compared with the BSE@$GW$. We hope that screened DDH can be the method of choice for evaluating the optical properties of chalcopyrite systems and different heterostructures where the BSE calculations are not feasible.

Finally, the overall quality
performances of screened-DDH are encouraging as they are quite close to many-body perturbation calculations in
providing the estimates of various properties. {\textcolor{black}{Importantly, the screened DDH is free of adjustable parameters (as opposed to HSE06).}} All the results are obtained by solving the generalized Kohn-Sham equation, and unlike the $GW$ method, it involves no virtual orbitals. This makes the functional easy to use with minimum computational cost. The present study shows this can be a method of choice for other chalcopyrite semiconductors, especially for Cu-based multinary semiconductors.


\section*{Acknowledgments}
SJ would like to thank Dr. Lucian A. Constantin for valuable
comments, suggestions, and technical details.


\twocolumngrid
\bibliography{reference.bib}
\bibliographystyle{apsrev.bst}

\end{document}